\documentclass[conference]{IEEEtran}
\usepackage{cite}
\usepackage{amsmath,amssymb,amsfonts}
\usepackage{algorithmic}
\usepackage{graphicx}
\usepackage{textcomp}
\usepackage{xcolor}
\usepackage{url}
\usepackage{stackengine}
\usepackage{subcaption}
\usepackage{makecell}
\usepackage{multirow}
\usepackage[hidelinks]{hyperref}
\usepackage{array}
\usepackage{balance}

\newcommand{\squishlist}{
	\begin{list}{$\bullet$}
		{ \setlength{\itemsep}{1pt}
			\setlength{\parsep}{1pt}
			\setlength{\topsep}{2.5pt}
			\setlength{\partopsep}{0.5pt}
			\setlength{\leftmargin}{1em}
			\setlength{\labelwidth}{1em}
			\setlength{\labelsep}{0.6em}
		}
	}
	\newcommand{\squishend}{
	\end{list}
}

\def\BibTeX{{\rm B\kern-.05em{\sc i\kern-.025em b}\kern-.08em
    T\kern-.1667em\lower.7ex\hbox{E}\kern-.125emX}}
\begin{document}

\title{Directory-Aware Query and Maintenance \\in Vector Databases
}

\author{
\IEEEauthorblockN{Mengzhao Wang\IEEEauthorrefmark{1}\IEEEauthorrefmark{4},
Zheng Gong\IEEEauthorrefmark{2},
Jingpei Hu\IEEEauthorrefmark{3},
Jiajie Fu\IEEEauthorrefmark{3},
Maojia Sheng\IEEEauthorrefmark{3},
Junwen Chen\IEEEauthorrefmark{3},
Yifan Zhu\IEEEauthorrefmark{2}
}
\IEEEauthorblockA{
\IEEEauthorrefmark{1}Hangzhou Dianzi University
\IEEEauthorrefmark{2}Zhejiang University
\IEEEauthorrefmark{3}ByteDance
}
\IEEEauthorblockA{wmzssy@hdu.edu.cn, \{gongzheng\_kurt, xtf\_z\}@zju.edu.cn,\\ \{hujingpei, fujiajie.168, shengmaojia, chenjunwen\}@bytedance.com}
}

\maketitle

\begin{abstract}
Vector databases typically manage metadata as flat scalar attributes, which limits their ability to express hierarchical directory semantics commonly used to organize code repositories, enterprise documents, and agent memories. As a result, directory-scoped retrieval and structural updates are often implemented as application-layer workarounds, making recursive scope resolution expensive and directory maintenance difficult to keep consistent. This paper studies native directory semantics as a first-class capability for vector databases. We formalize two core operators: Directory-Semantic Query (DSQ) for hierarchically scoped retrieval, and Directory-Semantic Maintenance (DSM) for structural updates. We then evaluate three implementation strategies: query-time path expansion (\textsc{PE-Online}), ingestion-time path expansion (\textsc{PE-Offline}), and a Trie-based Hierarchical Index (\textsc{TrieHI}). Our analysis exposes the fundamental limitations of expansion-based designs: flattening the hierarchy incurs high recursive-query latency in \textsc{PE-Online} and unscalable write amplification during structural changes in both expansion strategies. In contrast, \textsc{TrieHI} keeps the directory topology as a native prefix tree, enabling efficient recursive retrieval through tree traversal and reducing maintenance cost through topological node manipulation. We benchmark these design points within ByteDance's \textsf{Viking} vector search engine and release two large-scale datasets\footnote{\url{https://github.com/KurtPatrickHere/dir-vector-dataset} \label{dataset}}, WIKI-Dir and ARXIV-Dir, to support future research on directory-semantic vector search. Finally, \textsc{TrieHI} has been integrated into \textsf{OpenViking}\footnote{\url{https://github.com/volcengine/OpenViking} \label{openviking}}, an open-source context database for AI agents, where it supports filesystem-style context organization and directory-recursive retrieval.

\renewcommand{\thefootnote}{}
\footnotetext{\IEEEauthorrefmark{4}Work done while working with ByteDance.}
\end{abstract}

\vspace{-0.15cm}
\section{Introduction}
\label{sec-intro}
Vector databases~\cite{Survey-VDB-VLDBJ24,SingleStore-V,AlayaDB,HAKES,MicroNN,VSAG} have become a core data-management substrate for Retrieval-Augmented Generation (RAG) and AI-agent systems. They allow Large Language Models (LLMs) to retrieve external context from large corpora, including enterprise knowledge bases~\cite{Enterprise-KB}, public article archives such as Wikipedia~\cite{Wiki-WebSearch} and arXiv~\cite{arXiv-VDB}, and code repositories~\cite{CodeSearch}. Since generation quality depends heavily on the relevance and consistency of retrieved context~\cite{Chat2Data,LLM-DB-Survey,DB-Debug-RAG,MQA}, the basic vector-database pipeline---embedding data, indexing vectors, and executing $k$-nearest neighbor search---now sits on the critical path of systems such as Claude Code~\cite{claudecode}, OpenClaw~\cite{openclaw}, and Hermes~\cite{hermes}.

As these systems move from document collections to operational knowledge bases, retrieval requires more than nearest-neighbor similarity. The valid search space is often determined by structural scope: a project directory, a versioned documentation branch, a user-memory namespace, or an agent-skill repository. Current vector databases address part of this need through hybrid search and metadata filtering~\cite{HS-Milvus,HS-turbopuffer,HS-pinecone,HS-weaviate}. These mechanisms combine dense retrieval with sparse keyword matching~\cite{BM25,HS-FusionFunctions,HS-HybridIndex,OneSparse} and scalar predicates~\cite{VBase,HS-Apple,ADBV}, such as category = ``finance'' or $\text{price} < 100$. They are effective for flat attributes, but they do not model the hierarchical topology that defines many real retrieval scopes.

\begin{figure}
  \setlength{\abovecaptionskip}{0.1cm}
  \setlength{\belowcaptionskip}{-0.1cm}
  \centering
  \footnotesize
  \stackunder[2pt]{\includegraphics[scale=0.12]{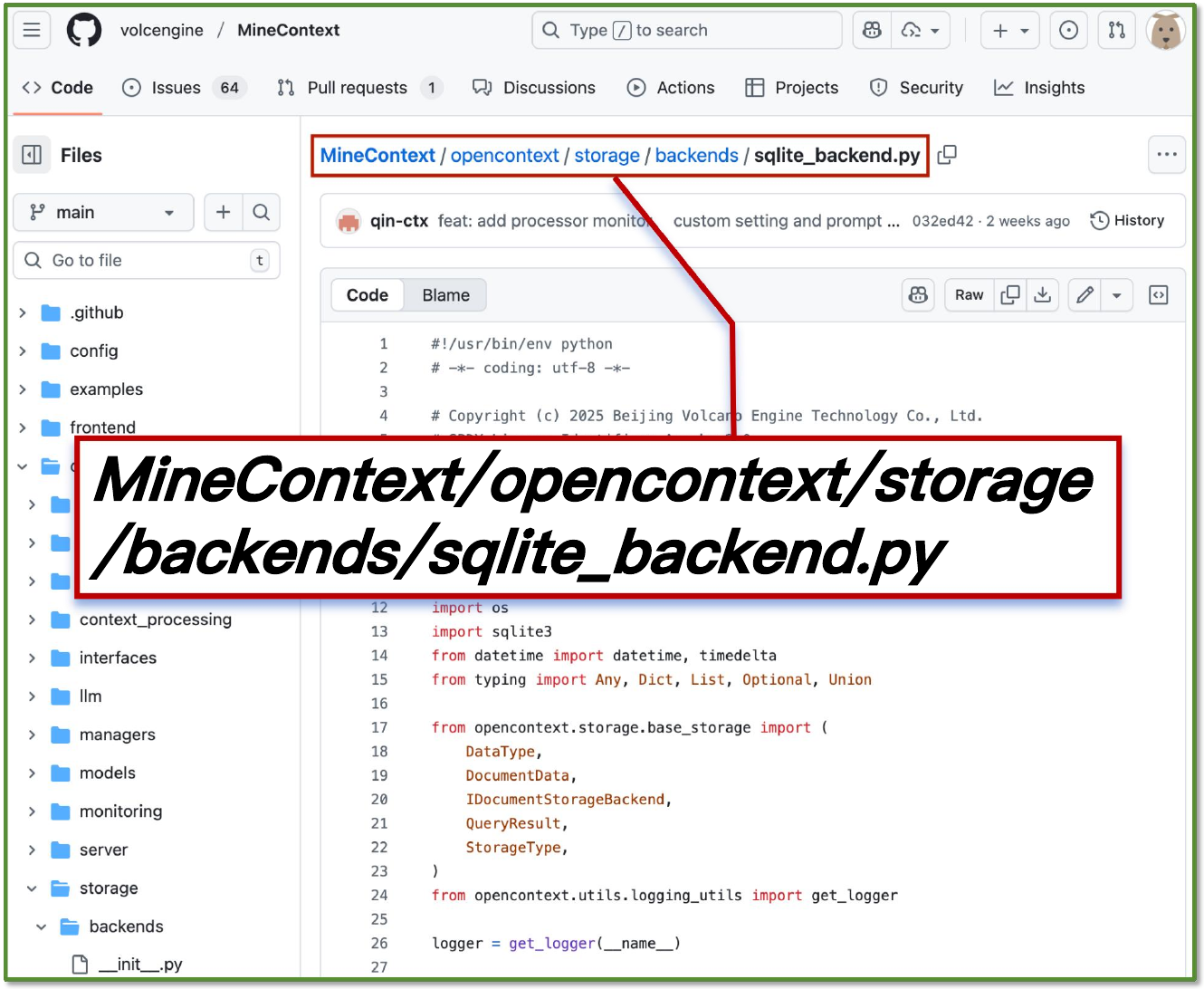}}{\hspace{0.1cm}(a) Code Repositories}
  % \hspace{0.15cm}
  \stackunder[2pt]{\includegraphics[scale=0.12]{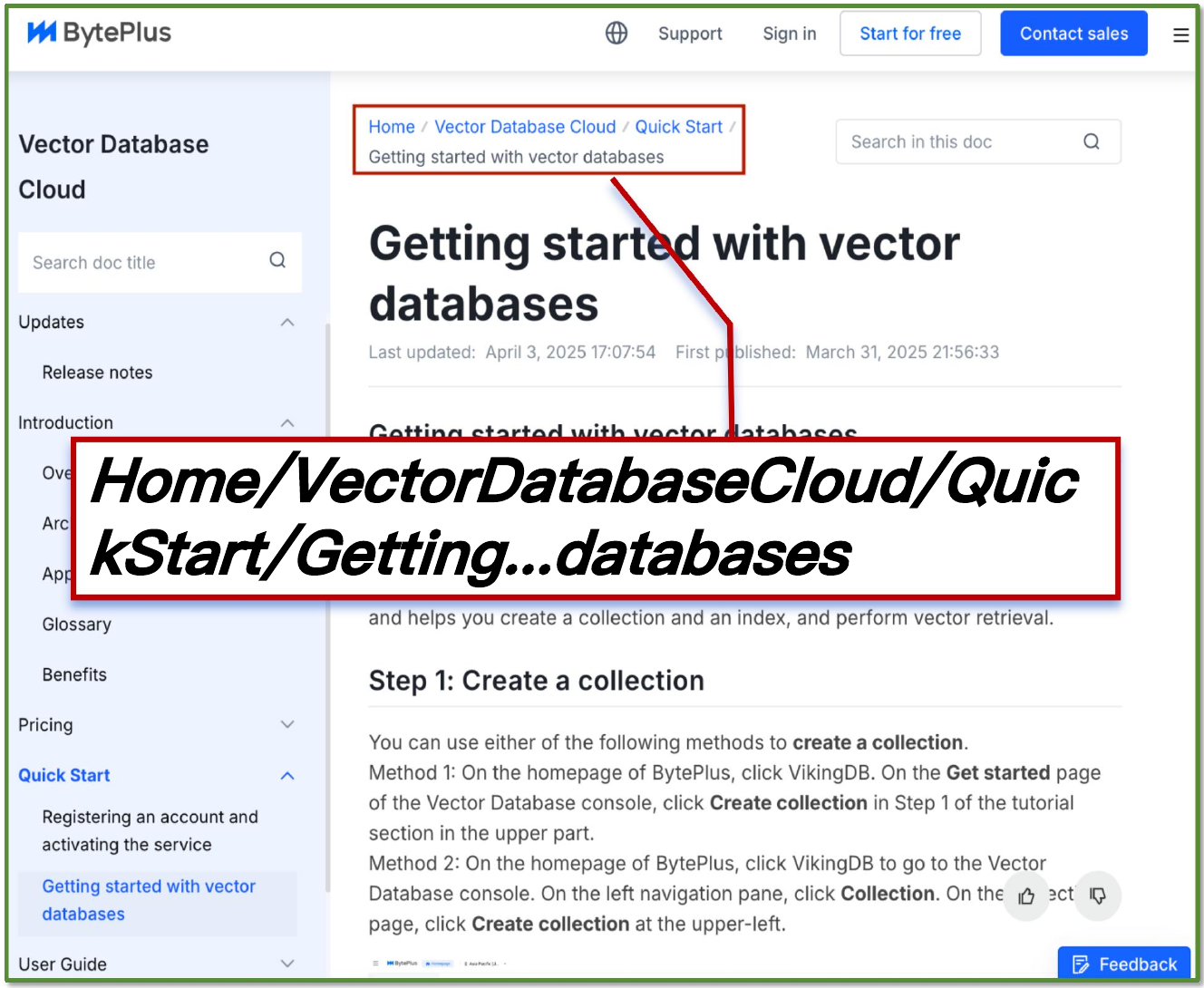}}{\hspace{0.1cm}(b) Corporate Wikis}
  % \hspace{0.15cm}
  \stackunder[2pt]{\includegraphics[scale=0.12]{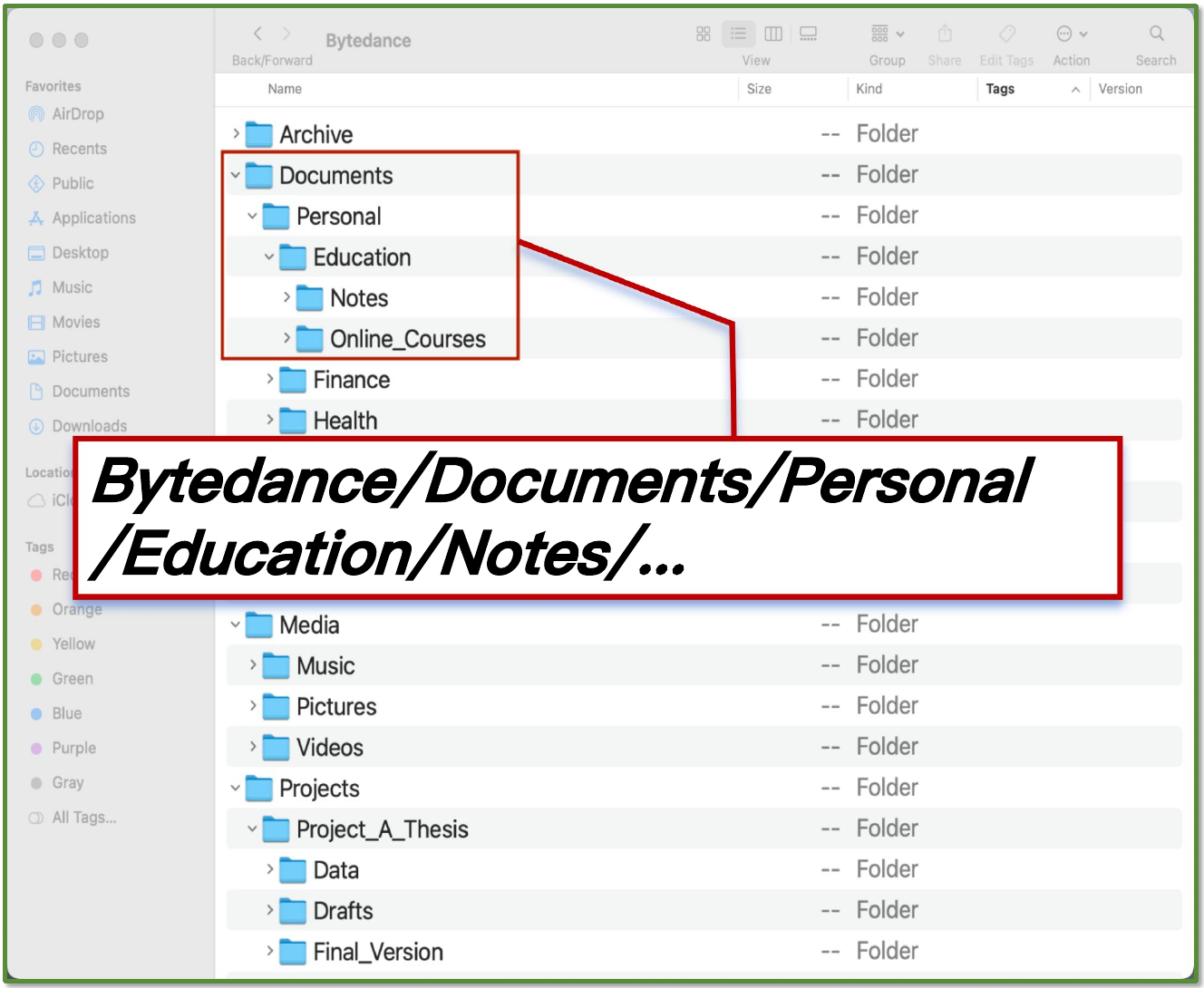}}{\hspace{0.1cm}(c) File Systems}
  \newline
  \caption{Knowledge bases with hierarchical directory structures.}
  \label{fig:dir_example}
  \vspace{-0.3cm}
\end{figure}

Directory structures are the most common such topology. Code repositories, corporate wikis, personal file systems, and agent context stores are all organized as nested namespaces, as illustrated in Figure~\ref{fig:dir_example}. In these settings, a path is not merely a string-valued attribute. A file such as \texttt{/src/core/payment.py} inherits meaning from its ancestors (e.g., \texttt{/src/} and \texttt{/src/core/}), and a query scoped to a directory must usually include or exclude entire subtrees. Flattening paths into scalar metadata discards this topology. The result is a recurring mismatch between semantic similarity and contextual validity: a query for ``architecture design'' may retrieve both current documents under \texttt{/docs/v2.0/} and obsolete documents under \texttt{/archive/v1.0/}, while relevant material under \texttt{/docs/} may be diluted by semantically similar but irrelevant files under \texttt{/logs/}. These failures are not ranking artifacts alone; they arise because the database lacks native operators for hierarchical scope.

Today, directory semantics are usually implemented outside the vector database. To search recursively under \texttt{/src/}, an application first enumerates descendant paths from an external metadata store and then submits a vector query with a large disjunctive predicate, such as \texttt{path IN [``/src/a/'', \dots]}. This workaround has two consequences. On the read path, recursive scope resolution becomes an opaque pre-processing step, preventing the database from co-optimizing structural pruning with vector search and often inflating latency. On the write path, directory mutations such as \texttt{MOVE} or \texttt{MERGE} require coordinated updates across an external directory store and vector metadata. A logical subtree operation degenerates into many document- or path-level updates, causing write amplification and exposing consistency hazards. These issues stem from keeping directory structure outside the vector database.

This paper argues that directory semantics should be first-class operations in vector databases. We formalize two primitives. Directory-Semantic Query (DSQ) scopes vector search by directory topology, supporting recursive subtree confinement, non-recursive directory access, and branch exclusion. Directory-Semantic Maintenance (DSM) captures structural mutations, including moving and merging subtrees. Together, DSQ and DSM provide the basic interface for expressing directory-semantic operations over vector data.

We study this design space by building a native directory-semantic module in ByteDance's \textsf{Viking} vector search engine. The module is decoupled from the underlying Approximate Nearest Neighbor (ANN) index, allowing directory semantics to be evaluated as a pluggable scope-resolution layer. We implement and compare three strategies: Online Path Expansion (\textsc{PE-Online}), Offline Path Expansion (\textsc{PE-Offline}), and a native Trie-based Hierarchical Index (\textsc{TrieHI}). The expansion-based strategies expose the limits of flattening: for recursive queries, \textsc{PE-Online} pays high query-time expansion cost, while \textsc{PE-Offline} shifts the cost to ingestion. Both remain vulnerable to subtree-wide maintenance amplification. \textsc{TrieHI} instead stores the hierarchy as a prefix tree, enabling recursive scope resolution through tree traversal and structural maintenance through tree-local operations.

\textsc{TrieHI} has also been integrated into \textsf{OpenViking}\textsuperscript{\ref{openviking}}, an open-source context database for AI agents. In \textsf{OpenViking}, \textsc{TrieHI} serves as the directory substrate for two core context-management functions: organizing memories, resources, and skills, and supporting directory-recursive retrieval that first locates promising directories and then searches their relevant descendants. Section~\ref{sec:openviking-integration} describes this integration, and Section~\ref{sec:openviking-case} evaluates its effect on agent-memory and knowledge-base Question Answering (QA) workloads.

Our main contributions can be summarized as follows:

\squishlist

\item We formulate directory semantics as a first-class vector-database abstraction. We define DSQ and DSM to capture hierarchy-scoped retrieval and topology-preserving maintenance, clarifying the requirements that cannot be represented by flat metadata filters alone.

\item We develop a pluggable directory-semantic layer in vector databases and instantiate three designs: \textsc{PE-Online}, \textsc{PE-Offline}, and \textsc{TrieHI}. The first two capture the natural expansion-based baselines, while \textsc{TrieHI} preserves the directory topology in a native prefix tree. We analyze their costs and trade-offs across query processing, storage, and structural maintenance.

\item We construct and release WIKI-Dir and ARXIV-Dir, two large-scale benchmarks with hierarchical directory topologies\textsuperscript{\ref{dataset}}. Using these datasets, we evaluate the three designs under recursive/non-recursive DSQ, DSM operations, and indexing/storage overheads with both graph-based and partition-based vector indexes.

\item We integrate \textsc{TrieHI} into the open-source \textsf{OpenViking} context database, where it supports filesystem-style organization over memories, resources, and skills, as well as directory-recursive retrieval. We evaluate this integration on user-memory and knowledge-base QA workloads.

\squishend

The rest of the paper is organized as follows. Section~\ref{sec-bg} formalizes directory-semantic operations. Sections~\ref{sec:expansion} and~\ref{sec:trie} present the expansion-based designs and \textsc{TrieHI}, respectively. Section~\ref{sec:experiment} reports the experimental results. Section~\ref{sec:related} reviews related work, and Section~\ref{sec:conclusion} concludes the paper.

\section{Background}
\label{sec-bg}

\subsection{Vector Database Execution Model}
Vector databases manage vectorized entries and execute similarity search through ANN indexes. A query vector is evaluated against an index such as a proximity graph (PG) or an inverted file (IVF), and the system returns the top-$k$ entries by vector similarity. Modern vector databases~\cite{2021milvus,MicroNN,opensearch} also support metadata filtering, a scalar predicate is resolved into a valid entry set, and the vector executor ranks only candidates that satisfy the predicate.

This execution model is a natural substrate for directory semantics. A directory constraint can be viewed as a scope predicate that determines entries eligible for vector ranking. The key difference from ordinary scalar filtering is that the predicate is topological rather than flat: resolving a path may require traversing ancestors, descendants, or sibling branches. Our goal is therefore not to replace ANN indexes, but to provide a directory-semantic scope-resolution layer that can feed valid candidate sets to existing vector-search executors. In our implementation, this layer is integrated into \textsf{Viking}, the vector search engine underlying \textsf{VikingDB}~\cite{byteplus-intro,byteplus}, which provides the PG and IVF indexes used in our experiments.

\subsection{Directory-Structured Knowledge Bases}
Hierarchical directories are a common organization model for human-generated and agent-managed knowledge~\cite{lynch2025directories,JacobsonKSS98,LutzRBD14}. A path is an ordered sequence of segments, and each prefix denotes a directory scope. For example, \texttt{/src/core/payment.py} is not merely a string identifier: the prefixes \texttt{/src/} and \texttt{/src/core/} encode structural context that distinguishes the entry from similarly named items in other branches.
Figure~\ref{fig:running_example} shows a simplified enterprise knowledge base used as a running example. The same abstraction applies to code repositories, file systems, and agent context stores. Each leaf entry has a vector payload, while each internal directory defines a scope over all entries in its subtree. A directory-aware vector database must therefore support both retrieval over such scopes and maintenance of the underlying topology as the namespace evolves.

\subsection{Directory-Semantic Operations}
We define two classes of operations over directory-structured vector data.

\squishlist
    \item \textit{Directory-Semantic Query (DSQ):} read-path operations that convert directory constraints into valid candidate scopes for vector search.
    \item \textit{Directory-Semantic Maintenance (DSM):} write-path operations that mutate the directory topology while preserving the intended namespace semantics for subsequent DSQs.
\squishend

\begin{figure}[t]
\centering
\includegraphics[width=0.78\linewidth]{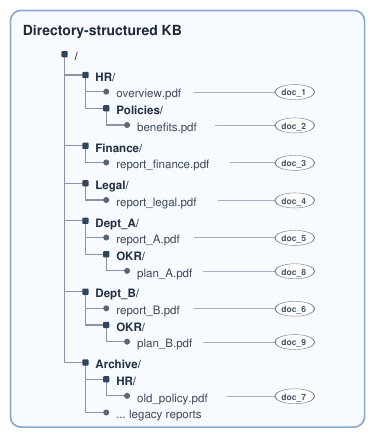}
\vspace{-0.1cm}
\caption{Example of a directory-structured knowledge base.}
\label{fig:running_example}
\vspace{-0.4cm}
\end{figure}

\vspace{0.1cm}
\noindent\textbf{DSQ Primitives.}
We focus on two base query primitives. A \textit{recursive query} searches all entries under a target directory, including entries in descendant directories. For example, a query anchored at \texttt{/HR/} considers both \texttt{doc\_1} and \texttt{doc\_2}; a query anchored at \texttt{/HR/Policies/} considers \texttt{doc\_2} while excluding relevant entries outside that subtree, such as \texttt{doc\_7} under \texttt{/Archive/HR/}. A \textit{non-recursive query} searches only entries directly bound to the target directory. For instance, a non-recursive query over \texttt{/HR/} includes \texttt{doc\_1} but excludes \texttt{doc\_2}, which resides in a child directory.

These two primitives are sufficient to express common derived operations. An exclusion query can be represented by subtracting the recursive scope of a branch. Sibling and ancestor queries can be reduced to combinations of directory traversal and recursive/non-recursive scope resolution. Accordingly, our design and evaluation focus on the two base DSQ primitives as the core execution workload.

\vspace{0.1cm}
\noindent\textbf{DSM Primitives.}
We use two structural mutations to characterize maintenance cost. A \textit{directory move} relocates an entire subtree to a new parent. In Figure~\ref{fig:running_example}, moving \texttt{/Dept\_A/} under \texttt{/Dept\_B/} changes the namespace of \texttt{Dept\_A} and all descendants while preserving their internal structure. A \textit{directory merge} consolidates a source subtree into a target subtree and reconciles name conflicts recursively. For example, merging \texttt{/Dept\_A/} into \texttt{/Dept\_B/} transfers \texttt{doc\_5} into the target branch and recursively reconciles the overlapping \texttt{OKR} directories containing \texttt{doc\_8} and \texttt{doc\_9}.

\texttt{MOVE} and \texttt{MERGE} capture the main challenge in directory maintenance: a single logical namespace operation may affect many descendants. Other lifecycle operations, such as directory creation, deletion, and renaming, can be modeled as simpler variants of these primitives. We therefore use \texttt{MOVE} and \texttt{MERGE} as representative DSM workloads.

\subsection{Design Requirements}
A directory-semantic layer for vector databases must satisfy four requirements.
First, it should provide \textit{scope correctness}: DSQ must resolve the intended recursive or non-recursive directory scope before vector ranking.
Second, it should provide \textit{query efficiency}: resolving a directory scope should not require scanning or expanding a large subtree on every query.
Third, it should provide \textit{maintenance efficiency}: structural mutations should avoid rewriting every entry or materialized path in the affected subtree whenever possible.
Fourth, it should preserve \textit{ANN-index independence}: directory semantics should act as a scope-resolution module that can work with different vector indexes, rather than requiring changes to the ANN algorithm.
\begin{figure}
  \setlength{\abovecaptionskip}{0.2cm}
  \setlength{\belowcaptionskip}{-0.2cm}
  \centering
  \includegraphics[width=\linewidth]{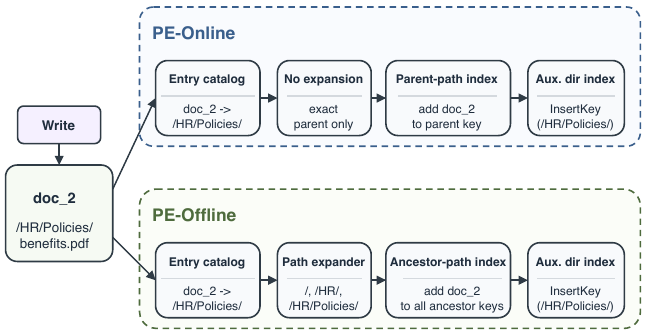}
  \caption{Comparison of ingestion workflows. \textsc{PE-Online} (top) indexes only the exact parent path; \textsc{PE-Offline} (bottom) pre-computes and indexes all ancestral paths.}
  \label{fig:expansion_ingest_path}
  \vspace{-0.4cm}
\end{figure}

\section{Path Expansion Strategies}
\label{sec:expansion}
\begin{figure}
  \setlength{\abovecaptionskip}{0.2cm}
  \setlength{\belowcaptionskip}{-0.2cm}
  \centering
  \includegraphics[width=.98\linewidth]{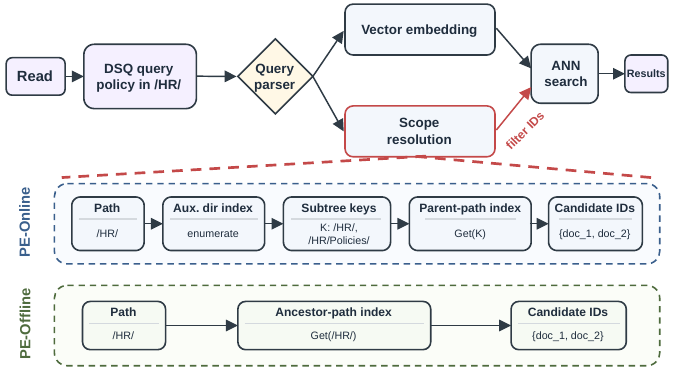}
  \caption{Recursive DSQ workflows. \textsc{PE-Online} (top) uses query-time path expansion and set union; \textsc{PE-Offline} (bottom) uses a single lookup for a pre-computed set.}
  \label{fig:expansion_query_path}
  \vspace{-0.3cm}
\end{figure}

Path expansion is the most direct way to implement directory semantics on top of a scalar-filtering vector database. The hierarchy is represented by path strings, and a directory constraint is rewritten into one or more scalar path predicates. For example, a recursive DSQ anchored at \texttt{/HR/} can be expanded into the path keys \texttt{/HR/} and \texttt{/HR/Policies/}.
This strategy keeps the ANN index unchanged and places all directory-specific logic in a metadata scope-resolution layer. The central design choice is when to expand the directory scope: at query time, by enumerating descendant paths, or at ingestion time, by materializing ancestor memberships. We study these two choices as Online Path Expansion (\textsc{PE-Online}) and Offline Path Expansion (\textsc{PE-Offline}). Figures~\ref{fig:expansion_ingest_path} and~\ref{fig:expansion_query_path} illustrate the corresponding ingestion and recursive-query workflows using the running example in Figure~\ref{fig:running_example}.

\subsection{Online Path Expansion (\textsc{PE-Online})}
\label{sec:pe-online}
\textsc{PE-Online} materializes only exact directory membership and defers recursive expansion to query time. It is therefore a \textit{time-for-space} design: ingestion and storage remain lightweight, but recursive queries must dynamically enumerate and combine the paths in the target subtree.

\vspace{0.1cm}
\noindent\textbf{Storage and Indexing.}
The metadata layer maintains three structures. First, an entry-to-directory mapping records the logical parent directory of each vectorized entry. Second, a parent-path inverted index maps each directory path key to the set of entries stored directly under that directory. Third, an auxiliary directory index stores all directory path keys and supports prefix enumeration and direct-child lookup. These structures are independent of the ANN index; they only compute the valid candidate set supplied to vector search.

As shown in Figure~\ref{fig:expansion_ingest_path} (top), when \texttt{doc\_2} is inserted under \texttt{/HR/Policies/}, \textsc{PE-Online} records its parent directory, inserts \texttt{doc\_2} only into the parent-path posting list for \texttt{/HR/Policies/}, and registers the path key in the auxiliary directory index if it is new. No ancestor posting lists are updated. Thus, each entry contributes one path posting.

\vspace{0.1cm}
\noindent\textbf{DSQ.}
A DSQ first resolves the directory constraint into a candidate entry set, and the ANN executor then ranks vectors within that set. In \textsc{PE-Online}, the resolution procedure depends on whether the query is recursive.

\underline{\textit{Recursive Query.}}
For a recursive DSQ anchored at path $p$, \textsc{PE-Online} enumerates all directory keys in the subtree of $p$ using the auxiliary directory index. In the running example, a recursive DSQ over \texttt{/HR/} returns the key set $K=\{\texttt{``/HR/''}, \texttt{``/HR/Policies/''}\}$. The system then retrieves the path posting list for each key in $K$ and unions the resulting sets, yielding Set\{\texttt{doc\_1}, \texttt{doc\_2}\}. The recursive scope of $p$ is the union of path sets over all directory keys in the subtree of $p$.

Let $m_q$ denote the number of directory keys returned by this subtree enumeration. The metadata work is linear in $m_q$: one enumeration step that returns $m_q$ keys, up to $m_q$ path lookups, and a union over the returned posting lists. Thus, \textsc{PE-Online} is sensitive to broad or shallow directory scopes.

\underline{\textit{Non-Recursive Query.}}
A non-recursive DSQ over path $p$ does not expand the subtree. It directly looks up the path posting list for $p$. For example, a non-recursive DSQ over \texttt{/HR/} returns Set\{\texttt{doc\_1}\} and excludes \texttt{doc\_2} because \texttt{doc\_2} is stored under the child directory \texttt{/HR/Policies/}. This operation requires one path-key lookup plus the cost of reading the corresponding posting list.

\vspace{0.1cm}
\noindent\textbf{DSM.}
DSM operations in \textsc{PE-Online} update the path-key space maintained by the auxiliary and parent-path inverted indexes. The important point is that the update is performed at the \textit{directory-key} level: entries remain attached to their logical parent directories, but the scalar path keys used by the metadata indexes must be remapped or merged when the namespace changes. Therefore, the cost scales with the number of affected directory keys, not necessarily with the number of vectorized entries in the subtree.

\underline{\textit{Directory Move.}}
Consider moving \texttt{/Dept\_A/} under \texttt{/Dept\_B/}. \textsc{PE-Online} first enumerates the affected source keys, such as \texttt{/Dept\_A/} and \texttt{/Dept\_A/OKR/}, from the auxiliary directory index. It then computes the corresponding target keys by replacing the source prefix with the destination prefix. For each source-target key pair, the parent-path posting list is remapped from the old key to the new key, and the auxiliary directory index is updated by deleting the old key and inserting the new one. After these updates, future DSQs resolve the moved subtree through the new path keys.

The dominant metadata cost is $O(m_u)$, where $m_u$ is the number of directory keys in the moved subtree. This includes enumerating the affected keys, computing their target names, remapping parent-path posting lists, and updating the auxiliary directory index.

\underline{\textit{Directory Merge.}}
To merge \texttt{/Dept\_A/} into \texttt{/Dept\_B/}, \textsc{PE-Online} enumerates all source keys under \texttt{/Dept\_A/} and computes their target keys under \texttt{/Dept\_B/}. For each source-target pair, the system retrieves the source posting list and the existing target posting list, unions them when they share the same path name, writes the merged posting list under the target key, and removes the source key. In the running example, the source key \texttt{/Dept\_A/OKR/} is merged into the target key \texttt{/Dept\_B/OKR/}, combining \{\texttt{doc\_8}\} and \{\texttt{doc\_9}\}.

The merge also performs $O(m_u)$ metadata updates over source directory keys, with additional set-union work for target-key conflicts. This makes \texttt{MERGE} no cheaper than \texttt{MOVE}, and potentially more expensive when many source paths collide with existing target paths.

\begin{table*}[t]
\centering
\caption{Performance trade-offs of expansion-based designs. $m_q$ is the number of directory keys in the queried subtree, $m_u$ is the number of directory keys in the mutated subtree, $t$ is the relevant directory depth or affected-ancestor bound, and $c$ is the number of immediate child directories.}
\vspace{-0.2cm}
\label{tab:expansion_tradeoffs}
\begingroup
\scriptsize
\setlength{\tabcolsep}{5pt}
\renewcommand{\arraystretch}{1.22}
\begin{tabular*}{\textwidth}{@{\extracolsep{\fill}}>{\raggedright\arraybackslash}m{0.15\textwidth}|m{0.39\textwidth}|m{0.39\textwidth}@{}}
\hline
\textbf{Aspect} & \textbf{\makecell{\textsc{PE-Online}\\(Query-Time Expansion)}} & \textbf{\makecell{\textsc{PE-Offline}\\(Ingestion-Time Expansion)}} \\ \hline
Index Storage & \textbf{Low}: one parent posting list per entry & \textbf{High}: $t$ ancestor posting lists per entry \\ \hline
Ingestion Work & \textbf{Low}: one posting update plus one auxiliary key update & \textbf{High}: $t$ posting updates plus one auxiliary key update \\ \hline
Recursive DSQ & \textbf{High metadata work}: subtree enumeration, up to $m_q$ posting list lookups, and set union & \textbf{Low metadata work}: one parent-path posting list lookup \\ \hline
Non-Recursive DSQ & \textbf{Low metadata work}: one parent-path posting list lookup & \textbf{High metadata work}: one parent-path posting list lookup, $c$ child-path posting list lookups, and set difference \\ \hline
\texttt{MOVE} & \textbf{High maintenance work}: $O(m_u)$ path-key remapping & \textbf{High maintenance work}: $O(m_u)$ path-key remapping plus $O(t)$ ancestor updates \\ \hline
\texttt{MERGE} & \textbf{High maintenance work}: $O(m_u)$ path-key merging, plus conflict unions & \textbf{High maintenance work}: $O(m_u)$ path-key merging plus $O(t)$ ancestor updates, plus conflict unions \\ \hline
\end{tabular*}
\endgroup
\vspace{-0.2cm}
\end{table*}

\subsection{Offline Path Expansion (\textsc{PE-Offline})}
\label{sec:pe-offline}
\textsc{PE-Offline} makes the opposite design choice. Instead of expanding a subtree at query time, it materializes ancestor memberships during ingestion. Each entry is indexed not only under its exact parent directory, but also under every ancestor directory. This is a \textit{space-for-time} design: recursive DSQ can be resolved by one path-key lookup, while ingestion, storage, non-recursive DSQ, and DSM become more expensive.

\vspace{0.1cm}
\noindent\textbf{Storage and Indexing.}
The metadata layer remains independent of the ANN index. It maintains: (1) an entry-to-directory mapping for the logical parent directory of each vectorized entry; (2) an auxiliary directory index over all directory path keys, used for direct-child lookup and DSM path-key enumeration; (3) a path expander that maps an exact parent path to its ancestor path sequence; and (4) an ancestor-materialized inverted index mapping each directory key to the set of entries located at or below that directory.

Figure~\ref{fig:expansion_ingest_path} (bottom) illustrates the ingestion of \texttt{doc\_2} under \texttt{/HR/Policies/}. The path expander generates the ancestor sequence [\texttt{``/''}, \texttt{``/HR/''}, \texttt{``/HR/Policies/''}], and \texttt{doc\_2} is inserted into the posting list of each key. The auxiliary directory index registers the exact parent path if it is new. Therefore, an entry at depth $t$ contributes $t$ ancestor postings. This replication is what makes recursive DSQ resolution cheap in metadata lookups, but it also means high-level directory keys, such as \texttt{``/''}, materialize large descendant sets.

\vspace{0.1cm}
\noindent\textbf{DSQ.}
As before, DSQ first resolves a directory constraint into a candidate entry set and then passes the set to the ANN executor. Because \textsc{PE-Offline} stores ancestor posting lists, recursive and non-recursive DSQ have opposite cost profiles.

\underline{\textit{Recursive Query.}}
A recursive DSQ over path $p$ is resolved by a single lookup of the ancestor posting list for $p$. For example, querying \texttt{/HR/} directly returns Set\{\texttt{doc\_1}, \texttt{doc\_2}\}, because both entries were inserted into the \texttt{/HR/} posting list at ingestion.

\underline{\textit{Non-Recursive Query.}}
For a non-recursive DSQ, the exact-parent entries under $p$ must be separated from descendant entries already materialized in $p$'s posting list. \textsc{PE-Offline} retrieves the posting list $Set_{Total}$ for $p$, enumerates the direct child directory keys of $p$ from the auxiliary directory index, retrieves their posting lists, unions them into $Set_{Children}$, and computes $Set_{Total} \setminus Set_{Children}$. For \texttt{/HR/}, this gives Set\{\texttt{doc\_1}, \texttt{doc\_2}\} $\setminus$ Set\{\texttt{doc\_2}\} = Set\{\texttt{doc\_1}\}. If $c$ is the number of direct child directories, scope resolution requires one lookup for $p$, one child enumeration, up to $c$ child lookups, and the corresponding union/difference operations.

\vspace{0.2cm}
\noindent\textbf{DSM.}
\textsc{PE-Offline} must maintain two kinds of metadata during DSM: the path keys for directories inside the mutated subtree, and the ancestor posting lists outside the subtree. This second requirement is the main difference from \textsc{PE-Online}. When a subtree changes parent, the aggregate entry set of the subtree must be removed from ancestor directories that no longer contain it and added to ancestor directories that newly contain it.

\underline{\textit{Directory Move.}}
\textsc{PE-Offline} handles a move, such as relocating \texttt{/Dept\_A/} under \texttt{/Dept\_B/}, in two coordinated steps. First, it enumerates the $m_u$ directory keys in the moved subtree and computes their new keys by prefix substitution. The posting list associated with each subtree key is remapped from the old key to the new key, and the auxiliary directory index is updated accordingly. Second, it updates ancestor memberships outside the moved subtree. Let $S$ be the aggregate posting list of the moved subtree root, and let $A^{-}$ and $A^{+}$ denote the old-only and new-only \textit{proper} ancestor sets after removing common proper ancestors. The system removes $S$ from posting lists in $A^{-}$ and adds $S$ to posting lists in $A^{+}$. Common proper ancestors are left unchanged because they contain the subtree before and after the move. The posting list of the moved subtree root is not updated through this ancestor-membership step; it is carried from the old root key to the new root key during path-key remapping.

The metadata cost has two components: $O(m_u)$ path-key remapping for the moved subtree and $O(t)$ ancestor-membership updates, where $t$ bounds the number of affected ancestors.

\underline{\textit{Directory Merge.}}
A merge, such as merging \texttt{/Dept\_A/} into \texttt{/Dept\_B/}, combines the source subtree with an existing target subtree. \textsc{PE-Offline} enumerates all source directory keys, computes their target keys, and processes each source-target pair. If the target key does not exist, the source materialized posting list can be moved to the target key. If the target key already exists, the source and target posting lists are unioned and stored under the target key, after which the source key is removed from metadata indexes. In the running example, \texttt{/Dept\_A/OKR/} maps to the existing \texttt{/Dept\_B/OKR/} key, so their posting lists are merged.

The ancestor-membership update follows the same rule as \texttt{MOVE}: the aggregate source set is removed from old-only proper ancestors and added to new-only proper ancestors, while common proper ancestors remain unchanged. The target root itself is updated by the source-target path-key merge described above. The total metadata work is $O(m_u+t)$ plus the cost of set unions for target-key conflicts. This makes \texttt{MERGE} at least as expensive as \texttt{MOVE} under path expansion, and often more expensive when many source keys collide with existing target keys.

\subsection{Analysis of Expansion-Based Solutions}
Table~\ref{tab:expansion_tradeoffs} summarizes the consequence of implementing directory semantics through scalar path expansion. The main advantage of this approach is compatibility: both variants can reuse a vector database's existing scalar-filtering interface without modifying the ANN index. However, because the hierarchy is encoded only as scalar path strings, the metadata layer has no native subtree object to reuse across DSQ and DSM. Directory membership must therefore be reconstructed at query time or pre-materialized at ingestion time.

\textsc{PE-Online} chooses query-time expansion. This keeps storage and ingestion compact, but makes recursive DSQ sensitive to the directory cardinality of the queried subtree. \textsc{PE-Offline} chooses ingestion-time expansion. This makes recursive DSQ cheap to resolve in the metadata layer, but introduces ancestor-level write amplification and turns non-recursive DSQ into a set-difference operation over direct child subtrees. Thus, the two schemes expose a clear latency--space--ingestion trade-off.
More importantly, both schemes inherit the same maintenance bottleneck. A structural mutation changes the scalar namespace used to encode a subtree, so the affected path-key range must be remapped or merged. Ancestor materialization further requires updates to proper ancestor posting lists in \textsc{PE-Offline}. Consequently, expansion-based designs can express directory semantics, but they do so through indirect metadata rewrites whose cost remains tied to the mutated subtree size.
\section{Trie-based Hierarchical Index}
\label{sec:trie}
We next describe the data structures and operations of \textsc{TrieHI}, analyze its operational profile, and then discuss its deployment in \textsf{OpenViking}.

\subsection{\textsc{TrieHI} Design}
\label{sec:trie-design}
Section~\ref{sec:expansion} shows that scalar path expansion can express directory semantics, but it treats the directory hierarchy as a collection of path keys. This representation is convenient for reuse of scalar filtering, yet it does not provide a native object for a directory subtree. As a result, recursive scope resolution must be supported either by query-time descendant enumeration in \textsc{PE-Online} or by ancestor materialization in \textsc{PE-Offline}, and structural updates still require path-key remapping or merging. Trie-based Hierarchical Index (\textsc{TrieHI}) takes a different physical design: it keeps the directory topology as a native prefix tree and uses the tree nodes as reusable scope objects for DSQ and DSM.

\textsc{TrieHI} remains a metadata index layered above the ANN executor. It does not change vector ranking or the layout of the underlying vector index. Given a directory constraint, it resolves the corresponding candidate entry-ID set and passes that set to the vector executor, following the execution model in Section~\ref{sec-bg}. Its main design goal is to replace subtree-wide path-key manipulation with tree-local navigation.

\begin{figure}[t]
\centering
\includegraphics[width=\linewidth]{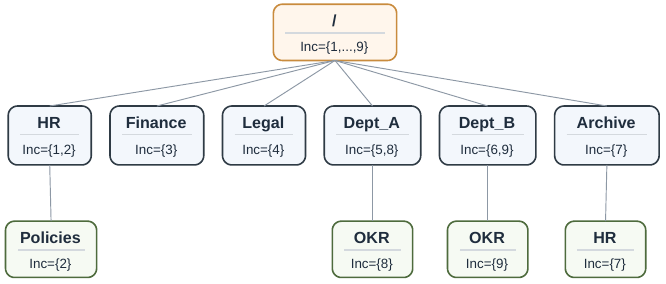}
\caption{The \textsc{TrieHI} structure for the running example.}
\label{fig:trie_structure}
\vspace{-0.4cm}
\end{figure}

\vspace{0.1cm}
\noindent\textbf{Storage and Indexing.}
\textsc{TrieHI} represents each directory as a \textit{TrieNode}. The root node corresponds to \texttt{/}, and each edge is labeled by one path segment. As illustrated in Figure~\ref{fig:trie_structure}, a node maintains the following fields: (1) \textit{segment}, the local directory name; (2) \textit{children}, a hash map from child segment names to child nodes; (3) \textit{parent}, a pointer to the parent node; and (4) \textit{inclusive\_ids}, the set of vectorized entry identifiers whose logical directory lies at the node or in one of its descendants. An entry identifier denotes the logical payload indexed by the vector database.

For notation, let $Inc(v)$ denote the \textit{inclusive\_ids} field of node $v$, let $Local(v)$ denote the entries directly bound to $v$, and let $Child(v)$ denote the immediate child nodes of $v$. The invariant maintained by \textsc{TrieHI} is:
\begin{equation}
\label{eq:trie_inclusive_invariant}
    Inc(v) =
    Local(v) \cup \bigcup_{w \in Child(v)} Inc(w).
\end{equation}
Thus, a node's inclusive set contains both entries directly stored at that directory and entries aggregated from its child subtrees. This invariant makes a directory node a reusable materialized scope. A recursive DSQ can use the aggregate set stored at the target node, while DSM operations update only the ancestor aggregates whose descendant membership changes. A separate entry-to-node catalog records the logical parent directory of each vectorized entry and supports namespace-level bookkeeping.

Ingestion follows the trie topology. To insert an entry such as \texttt{doc\_8} under \texttt{/Dept\_A/OKR/}, \textsc{TrieHI} first traverses the path from the root, creating missing nodes for \texttt{Dept\_A} or \texttt{OKR} if necessary. It then registers the entry with the terminal node and adds the entry ID to \textit{inclusive\_ids} along the terminal node and its ancestors. Thus, ingestion performs $O(t)$ node visits and $O(t)$ aggregate-set updates for a path of depth $t$. The cost is comparable in form to ancestor materialization in \textsc{PE-Offline}, but the hierarchy itself is represented as nodes and pointers rather than scalar path keys.

\vspace{0.1cm}
\noindent\textbf{DSQ.}
\textsc{TrieHI} resolves DSQ predicates by path traversal followed by set access or set difference.

\underline{\textit{Recursive Query.}}
For a recursive DSQ anchored at path $p$, \textsc{TrieHI} traverses the trie from the root to the node representing $p$. It then reads \textit{inclusive\_ids} from that node. For example, a recursive DSQ over \texttt{/Dept\_A/} resolves to Set\{\texttt{doc\_5}, \texttt{doc\_8}\} in the running example. The metadata access pattern is a path traversal of length $t$ plus one aggregate-set access; vector ranking is then performed within the resolved candidate set. Unlike \textsc{PE-Online}, no descendant path enumeration is required at query time.

\underline{\textit{Non-Recursive Query.}}
A non-recursive DSQ over path $p$ should include entries directly bound to $p$ but exclude entries in child directories. \textsc{TrieHI} computes this scope from the aggregate invariant. It retrieves $Set_{Total}=Inc(p)$, unions the $Inc(\cdot)$ sets of the immediate child nodes into $Set_{Children}$, and returns $Set_{Total}\setminus Set_{Children}$. For \texttt{/Dept\_A/}, this yields Set\{\texttt{doc\_5}, \texttt{doc\_8}\} $\setminus$ Set\{\texttt{doc\_8}\} = Set\{\texttt{doc\_5}\}. If $c$ is the number of immediate children, the metadata work consists of one path traversal, up to $c$ child-set accesses, and the corresponding union/difference operations.

\underline{\textit{Derived Path Patterns.}}
The prefix-tree representation also provides a natural execution substrate for derived path patterns, such as prefix-constrained or wildcard path segments. A wildcard segment can be evaluated by matching child names at the corresponding level and continuing traversal only along matching branches. This is a structural advantage over scanning flat path strings. Formalizing and evaluating these derived path predicates is a natural extension of the DSQ model and is left to future work.

\vspace{0.1cm}
\noindent\textbf{DSM.}
DSM operations in \textsc{TrieHI} update two states: the tree topology and the \textit{inclusive\_ids} aggregates required for future DSQs. The important difference from expansion-based designs is that a subtree has a stable node identity. A namespace mutation can therefore relink the subtree root instead of regenerating scalar path keys for every descendant directory. The aggregate sets still need to be updated along the ancestor chains whose descendant membership changes.

\underline{\textit{Directory Move.}}
Consider moving \texttt{/Dept\_A/} to \texttt{/Dept\_B/}. \textsc{TrieHI} locates the source node $s$, its old parent, and the new parent. Let $S=Inc(s)$ contain the entries directly bound to the moved root $s$ and its descendants. The system computes the old-only and new-only proper ancestor chains after removing common ancestors. It removes $S$ from \textit{inclusive\_ids} on old-only ancestors and adds $S$ to \textit{inclusive\_ids} on new-only ancestors. Finally, the source node is removed from the old parent's \textit{children} map, inserted into the new parent's \textit{children} map, and assigned the new parent pointer.

This operation visits nodes on the source and destination paths and updates aggregate sets on affected ancestors. It is independent of the number of descendant directory keys.

\underline{\textit{Directory Merge.}}
\texttt{MERGE} reconciles a source subtree with an existing target subtree. We describe the common case where the source and target roots are disjoint. Let $s$ be the source root and $d$ be the target root. \textsc{TrieHI} first uses $S=Inc(s)$ to update ancestor aggregates: $S$ is removed from old-only proper ancestors of $s$ and added to $d$ and the new-only proper ancestors of $d$, while common ancestors remain unchanged. It then reconciles the topology below $s$ and $d$. A non-conflicting source child can be relinked directly under the target node by updating one parent pointer and the corresponding child maps. If a source child and target child have the same directory name, \textsc{TrieHI} recursively merges that child pair.

The merge cost has a depth-dependent component for locating the source and target roots and updating their ancestor aggregates, plus a conflict-dependent component for recursively reconciling overlapping branches. Let $r$ denote the number of source nodes visited by this recursive reconciliation. In the worst case, $r$ can approach the number of directory nodes in the source subtree; in common cases with few name conflicts, non-conflicting child subtrees are relinked as whole units. The distinction from path expansion is that \textsc{TrieHI} performs conflict handling through node-level recursive merging: non-conflicting subtrees are relinked as units, and conflicting branches are reconciled on trie nodes rather than by remapping every descendant path key.

\underline{\textit{Consistency During Updates.}}
DSM updates in \textsc{TrieHI} preserve the aggregate invariant in Equation~\ref{eq:trie_inclusive_invariant} by applying each namespace mutation within its affected trie region. Before a mutation, \textsc{TrieHI} identifies that region: move covers the source subtree and destination path, and merge covers the source and target subtrees. Mutations on overlapping paths are serialized. Within one mutation, \textsc{TrieHI} follows a fixed order: collect the affected entry set, change the parent/child links, refresh the entry-to-node catalog and ancestor aggregates, and remove obsolete namespace bindings after relocation. After this sequence, topology, catalog entries, and \textit{inclusive\_ids} remain consistent. DSQ reads candidate sets from this metadata, while the ANN executor only receives the resolved set.

\begin{table}[t]
\centering
\caption{Operational profile of \textsc{TrieHI}: $t$ denotes directory depth, $c$ is the number of immediate child directories, and $r$ is the number of reconciled source nodes in \texttt{MERGE}.}
\vspace{-2mm}
\label{tab:trie_tradeoffs}
\scriptsize
\setlength{\tabcolsep}{4pt}
\renewcommand{\arraystretch}{1.18}
\begin{tabular*}{\columnwidth}{@{\extracolsep{\fill}}>{\raggedright\arraybackslash}m{0.26\columnwidth}|m{0.68\columnwidth}@{}}
\hline
\textbf{Aspect} & \textbf{\textsc{TrieHI}} \\ \hline
Index Storage & Trie topology plus per-node inclusive entry-ID sets \\ \hline
Ingestion Work & $O(t)$ node visits and aggregate-set updates \\ \hline
Recursive DSQ & $O(t)$ node visits plus one aggregate-set access \\ \hline
Non-Recursive DSQ & $O(t+c)$ metadata accesses plus union/difference \\ \hline
\texttt{MOVE} & $O(t)$ node visits and ancestor aggregate updates \\ \hline
\texttt{MERGE} & $O(t+r)$ node work plus conflict-dependent set unions \\ \hline
\end{tabular*}
\vspace{-2mm}
\end{table}

\subsection{Analysis of \textsc{TrieHI}}
\label{sec:trie-analysis}
Table~\ref{tab:trie_tradeoffs} summarizes the operational profile of \textsc{TrieHI}. Compared with scalar path expansion, the key change is that a directory subtree is represented by a node identity rather than by all materialized descendant path keys. This directly benefits recursive DSQ and \texttt{MOVE}: recursive DSQ resolves a node aggregate after path traversal, and \texttt{MOVE} relinks the subtree root while updating only affected ancestor aggregates.
The node-centric representation reduces subtree-wide path-key maintenance, but it introduces other metadata costs. Ingestion updates aggregates along the ancestor chain, and storage includes both the trie topology and per-node inclusive entry-ID sets. Non-recursive DSQ computes a local set difference between the target node and its immediate child subtrees. \texttt{MERGE} depends on the number of conflicting source nodes that require recursive reconciliation. Overall, these costs are tied to tree navigation, ancestor updates, and actual merge conflicts, rather than the expansion-based maintenance pattern of remapping every descendant path key in the affected subtree.

\subsection{Deployment in \textsf{OpenViking}}
\label{sec:openviking-integration}
\textsf{OpenViking}\textsuperscript{\ref{openviking}} is a context database for AI agents that organizes memories, resources, and skills through a virtual filesystem. It requires both frequent namespace maintenance and low-latency recursive retrieval. Memories are created, consolidated, archived, and attached to projects or skills, while resource subtrees may be reorganized as a workspace evolves. Queries often start from a directory scope and request coherent descendants, such as a project subtree or a user-memory branch. \textsc{PE-Online} would enumerate descendants on this retrieval path; \textsc{PE-Offline} avoids query-time enumeration but amplifies namespace maintenance. \textsc{TrieHI} fits this workload by representing directory scopes as reusable trie nodes and expressing context reorganization as subtree-local topology updates. We integrate \textsc{TrieHI} into this namespace to support directory-scoped context management and retrieval.

\begin{figure}[t]
\centering
\includegraphics[width=\linewidth]{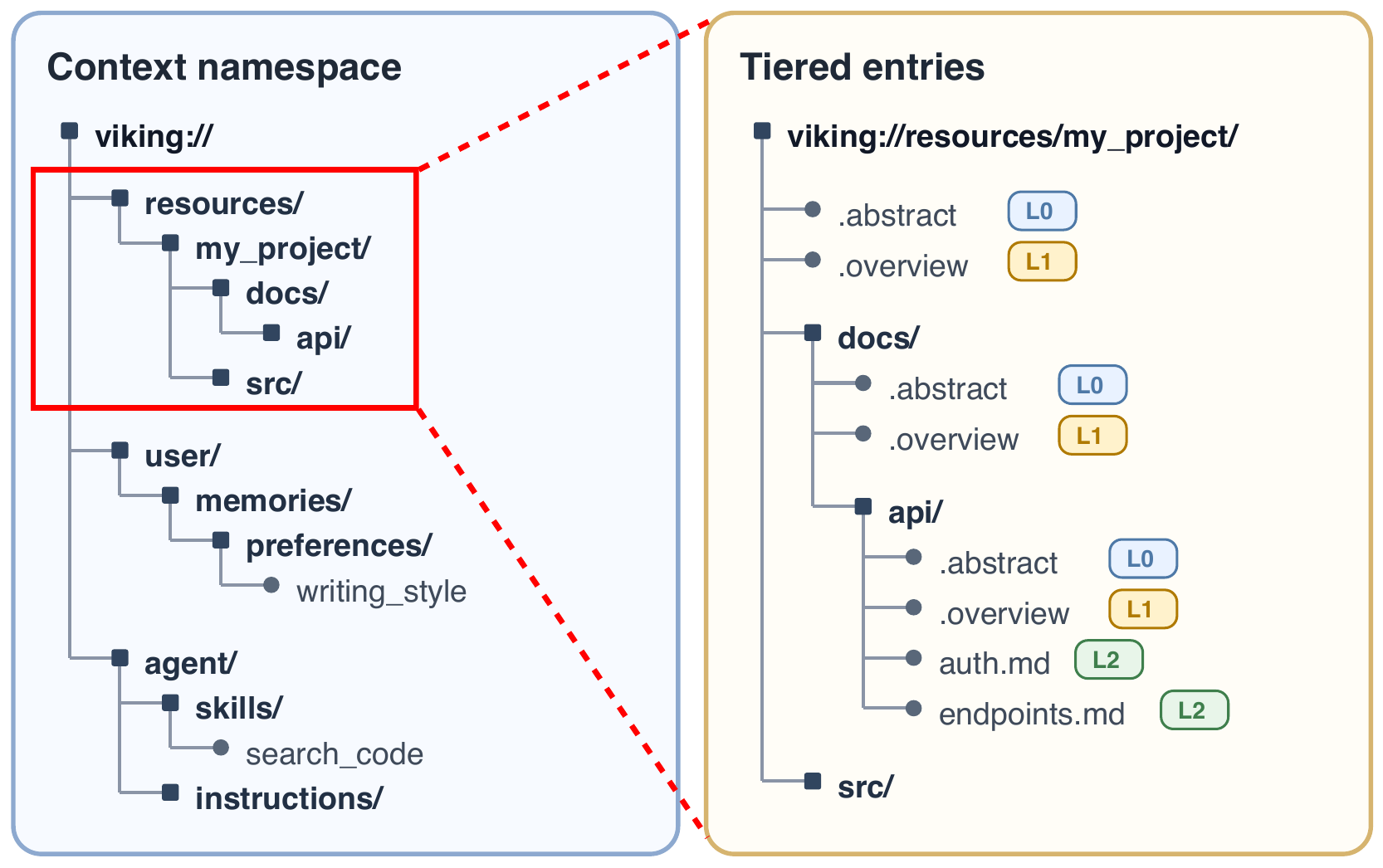}
\vspace{-0.2cm}
\caption{TrieHI-backed \textsf{OpenViking} namespaces. Left: filesystem-style context organization. Right: tiered context entries under the same hierarchy, where L0 denotes abstracts, L1 denotes overviews, and L2 denotes full content.}
\label{fig:openviking_namespace}
\vspace{-0.3cm}
\end{figure}

Figure~\ref{fig:openviking_namespace} illustrates two views of the same \textsc{TrieHI}-backed directory hierarchy in \textsf{OpenViking}. The first is a filesystem-style organization over different context types. The second stores tiered context entries: L0 abstracts support lightweight identification, L1 overviews expose directory-level summaries, and L2 entries keep the original content. We next discuss the three supported capabilities, summarized in Table~\ref{tab:openviking_integration}.

\vspace{0.10cm}
\noindent\textbf{Filesystem-Style Context Organization.}
\textsf{OpenViking} organizes memories, resources, and skills as \texttt{viking://} URIs in a hierarchical virtual filesystem. In the \textsc{TrieHI} abstraction, URI segments define directory nodes, and the same hierarchy serves both browsing and retrieval. For example, recursive context acquisition maps to recursive DSQ, while URI-subtree reorganization maps to DSM.

\vspace{0.10cm}
\noindent\textbf{Tiered Context Loading.}
\textsf{OpenViking} stores L0 abstracts, L1 overviews, and L2 full entries under shared directory scopes. In the \textsc{TrieHI} abstraction, these entries are associated with their directory scopes. A retrieval pipeline can therefore first operate on lightweight entries and then obtain detailed descendants when additional evidence is required.

\vspace{0.10cm}
\noindent\textbf{Directory-Recursive Retrieval.}
\textsf{OpenViking} supports retrieval over \texttt{viking://} directory scopes. Given a scoped query, \textsc{TrieHI} first resolves the requested directory as candidate entries; semantic vector scoring then ranks candidates within that scope for focused retrieval or aggregation. This realizes the directory scope needed to retrieve coherent surrounding context without scanning flat path strings.

\begin{table*}[t]
\centering
\caption{\textsf{OpenViking} capabilities realized by \textsc{TrieHI}.}
\vspace{-0.2cm}
\label{tab:openviking_integration}
\scriptsize
\setlength{\tabcolsep}{5pt}
\renewcommand{\arraystretch}{1.18}
\begin{tabular*}{\textwidth}{@{\extracolsep{\fill}}>{\raggedright\arraybackslash}m{0.23\textwidth}|m{0.37\textwidth}|m{0.32\textwidth}@{}}
\hline
\textbf{OpenViking capability} & \textbf{\textsc{TrieHI} role} & \textbf{Directory-semantic operation} \\ \hline
Filesystem-style organization & \texttt{viking://} URI segments are represented as directory nodes in the \textsc{TrieHI} abstraction. & Hierarchical namespace navigation; DSM for namespace changes. \\ \hline
Tiered context loading & L0/L1/L2 entries are associated with shared directory scopes in the \textsc{TrieHI} abstraction. & DSQ selects summary or full-detail entries under the same scope. \\ \hline
Directory-recursive retrieval & \textsc{TrieHI} resolves the requested directory scope and descendant candidates before vector ranking. & Recursive DSQ supplies scoped candidates for vector ranking and aggregation. \\ \hline
\end{tabular*}
\vspace{-0.2cm}
\end{table*}
\section{Experiments}
\label{sec:experiment}

This section evaluates the directory-semantic strategies introduced in Sections~\ref{sec:expansion} and~\ref{sec:trie}. The experiments are organized around four questions.

\squishlist
    \item \textbf{RQ1:} Can the designs resolve recursive and non-recursive directory scopes without dominating vector-search latency?
    \item \textbf{RQ2:} How do the designs behave when logical namespace mutations affect large subtrees?
    \item \textbf{RQ3:} What indexing-time and storage overheads are introduced by directory-semantic metadata?
    \item \textbf{RQ4:} When \textsc{TrieHI} is instantiated in \textsf{OpenViking}, do directory-scoped retrieval primitives improve realistic agent-memory and knowledge-base QA workloads?
\squishend
The first three questions are answered on WIKI-Dir and ARXIV-Dir, two directory-structured evaluation datasets constructed for DSQ and DSM. The last question is addressed by an end-to-end \textsf{OpenViking} evaluation, which measures the integrated context-database system.

\subsection{Experimental Setup}
\label{sec:exp-setup}

\noindent\textbf{Datasets.}
Since existing datasets do not expose directory-semantic operations, we construct two large-scale datasets\textsuperscript{\ref{dataset}}.

\underline{WIKI-Dir.}
WIKI-Dir is derived from a Wikipedia corpus with queries and relevance labels from DBpedia~\cite{DBPedia}. We add a hierarchical namespace by crawling the Wikipedia category hierarchy\footnote{\href{https://en.wikipedia.org/wiki/Category:Main_topic_articles}{\nolinkurl{https://en.wikipedia.org/wiki/Category:Main_topic_articles}}} for each article. Since Wikipedia categories form a directed acyclic graph, we construct a canonical tree by selecting one category path per entry and discarding invalid links. The resulting hierarchy contains 363,467 directories with an average depth of 11.95. The final dataset contains 1.94 million vectorized entries and 456 test queries with directory constraints at varying depths. Entries are encoded into 1024-dimensional vectors using \textit{bge-m3}~\cite{BGE-M3}. We also generate 1,000 \texttt{MOVE} operations and 1,000 \texttt{MERGE} operations for structural maintenance evaluation.

\underline{ARXIV-Dir.}
ARXIV-Dir is built from 2.76 million arXiv paper abstracts. Each entry is associated with two independent hierarchical namespaces derived from official arXiv metadata\footnote{\url{https://arxiv.org/category_taxonomy}}: a subject hierarchy, such as \texttt{/cs/AI/}, and a temporal hierarchy, such as \texttt{/2024/05/}. The combined namespace contains 168 subject directories with average depth 2.19 and 432 temporal directories with average depth 1.92. Abstracts are encoded into 1024-dimensional vectors using \textit{mxb-ai-embed-large-v1}~\cite{emb2024mxbai}. We sample 1,000 queries from the abstracts and generate directory constraints for each query. Ground truth is computed by brute-force search over the entries satisfying the corresponding constraints.

\vspace{0.12cm}
\noindent\textbf{Compared Designs.}
We compare the three designs analyzed in Sections~\ref{sec:expansion} and~\ref{sec:trie}.
\textsc{PE-Online} stores only the immediate path key and expands descendant paths at query time.
\textsc{PE-Offline} materializes ancestor memberships during ingestion so that recursive scopes can be resolved by a materialized path key.
\textsc{TrieHI} stores the hierarchy as a prefix tree and uses directory nodes as reusable scope objects.
All three designs are integrated with two ANN executors in the \textsf{Viking} engine: a graph-based proximity graph (PG) index and a partition-based inverted file (IVF) index. This lets us distinguish directory-semantic overhead from ANN-index behavior.

\vspace{0.12cm}
\noindent\textbf{Metrics.}
For DSQ, we report retrieval quality using nDCG@$k$ or Recall@$k$, together with end-to-end query latency. We also report directory-only latency, defined as the time required to resolve a directory predicate into a candidate entry-ID set before ANN ranking. For DSM, we measure wall-clock latency of \texttt{MOVE} and \texttt{MERGE}. For resource overhead, we report total index construction time and index size, including the baseline vector index and the directory-semantic module.

\vspace{0.12cm}
\noindent\textbf{Implementation Details.}
Entry-ID sets are represented by Roaring bitmaps~\cite{chambi2016better}, enabling compressed set union, intersection, and difference. The expansion-based methods use path-key metadata structures for scalar path lookup; \textsc{TrieHI} uses trie nodes and inclusive entry-ID sets as described in Section~\ref{sec:trie}. All methods maintain a common catalog that maps each entry to its current directory representation, such as a path key or a trie node, for maintenance. Because this catalog is required by every design, we exclude it when comparing DSM latency and directory-module indexing overhead; the reported differences therefore reflect the design-specific metadata structures and update procedures. The experiments run on a Linux server with a 48-core Intel(R) Xeon(R) Silver 4310 CPU @ 2.10GHz and 125GB RAM. The implementations are in C++ and integrated into ByteDance's \textsf{Viking} vector search engine.

\subsection{Directory-Semantic Query Performance}
\label{sec:query-perf}

We first evaluate the two base DSQ primitives: recursive and non-recursive scope queries.

\begin{figure}
  \setlength{\abovecaptionskip}{0cm}
  \setlength{\belowcaptionskip}{-0.1cm}
  \centering
  \footnotesize
  \stackunder[1pt]{\includegraphics[scale=0.28]{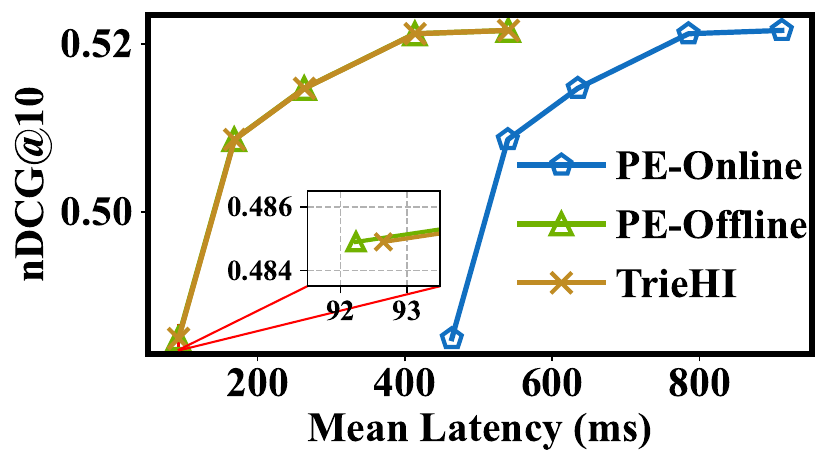}}{\hspace{0.1cm}(a) PG, WIKI-Dir}
  \stackunder[1pt]{\includegraphics[scale=0.28]{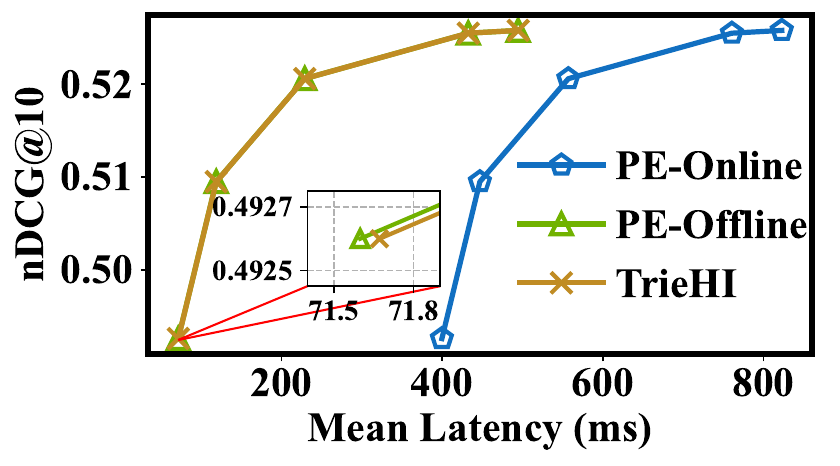}}{\hspace{0.1cm}(b) IVF, WIKI-Dir}
  \newline
  \stackunder[1pt]{\includegraphics[scale=0.28]{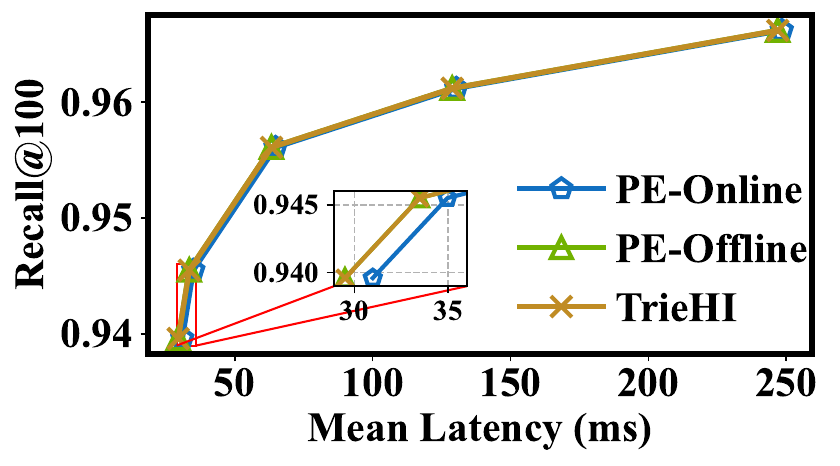}}{\hspace{0.1cm}(c) PG, ARXIV-Dir}
  \stackunder[1pt]{\includegraphics[scale=0.28]{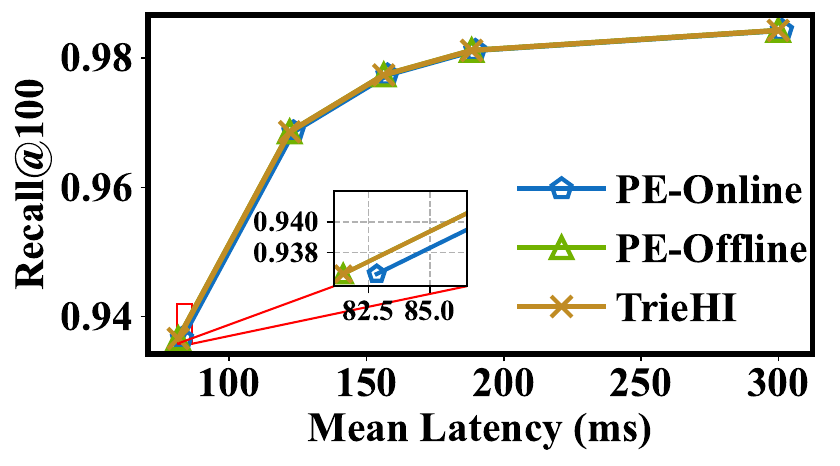}}{\hspace{0.1cm}(d) IVF, ARXIV-Dir}
  \newline
  \caption{Recursive DSQ performance: retrieval quality versus mean query latency.}
  \label{fig:recursive_query_performance}
  \vspace{-0.2cm}
\end{figure}

\begin{figure}
  \setlength{\abovecaptionskip}{0cm}
  \setlength{\belowcaptionskip}{-0.1cm}
  \centering
  \footnotesize
  \stackunder[1pt]{\includegraphics[scale=0.28]{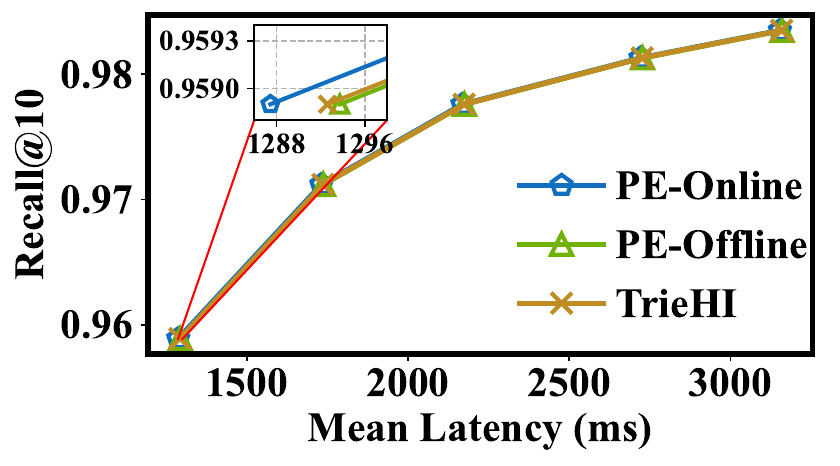}}{\hspace{0.2cm}(a) PG, WIKI-Dir}
  \stackunder[1pt]{\includegraphics[scale=0.28]{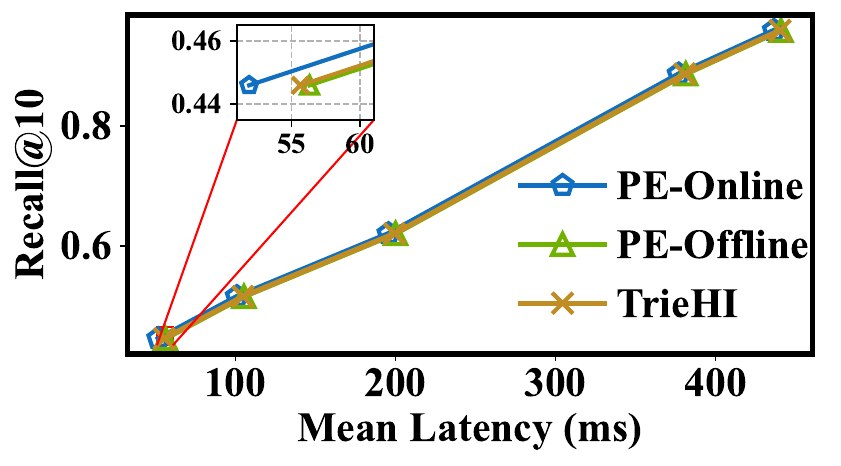}}{\hspace{0.2cm}(b) IVF, WIKI-Dir}
  \newline
  \stackunder[1pt]{\includegraphics[scale=0.28]{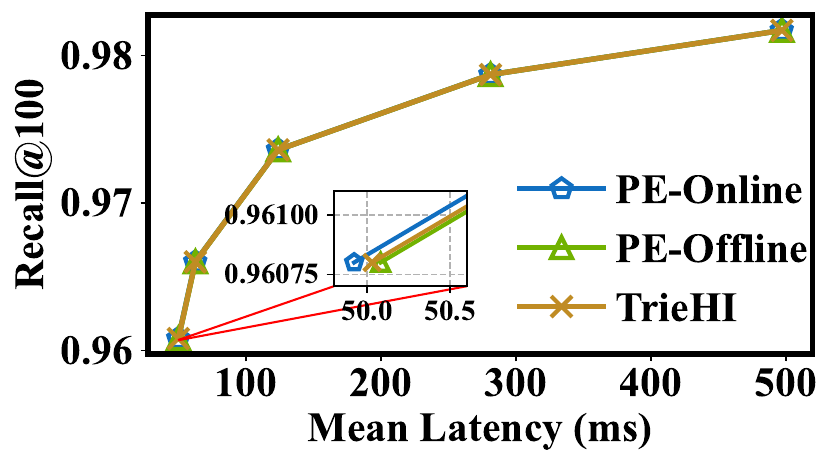}}{\hspace{0.2cm}(c) PG, ARXIV-Dir}
  \stackunder[1pt]{\includegraphics[scale=0.28]{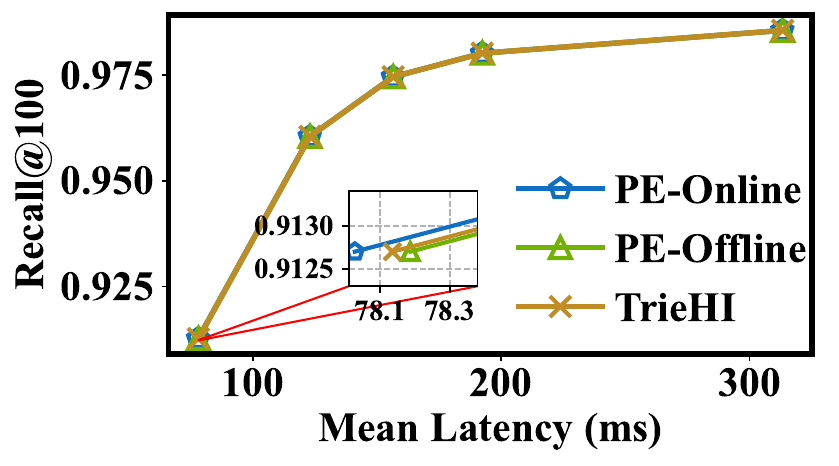}}{\hspace{0.2cm}(d) IVF, ARXIV-Dir}
  \newline
  \caption{Non-recursive DSQ performance: retrieval quality versus mean query latency.}
  \label{fig:non_recursive_query_performance}
  \vspace{-0.2cm}
\end{figure}

\begin{table*}[t!]
\centering
\caption{Directory-only latency ($\boldsymbol{\mu}$s) for candidate entry-ID set generation. The lowest latency in each group is bolded.}
\vspace{-0.2cm}
\label{tab:bitmap_latency}
\scriptsize
\setlength{\tabcolsep}{.0065\linewidth}
\renewcommand{\arraystretch}{1.16}
\begin{tabular*}{\textwidth}{@{\extracolsep{\fill}}>{\raggedright\arraybackslash}m{0.07\textwidth}|>{\raggedright\arraybackslash}m{0.09\textwidth}|ccccc|ccccc@{}}
\hline
\multirow{2}{*}{\textbf{\makecell{Query\\Type}}$\downarrow$} & \textbf{Dataset}$\rightarrow$ & \multicolumn{5}{c|}{\textbf{WIKI-Dir}} & \multicolumn{5}{c}{\textbf{ARXIV-Dir}} \\ \cline{2-12}
& \textbf{Method}$\downarrow$ & \textbf{Mean} & \textbf{P90} & \textbf{P95} & \textbf{P99} & \textbf{P99.9} & \textbf{Mean} & \textbf{P90} & \textbf{P95} & \textbf{P99} & \textbf{P99.9} \\ \hline
\multirow{3}{*}{\textbf{Recur.}} & \textsc{PE-Online} & 365,000.80 & 669,527.67 & 748,949.23 & 913,704.33 & 2,654,322.00 & 1,089.20 & 5,241.70 & 5,261.33 & 5,292.37 & 5,380.60 \\ \cline{2-12}
& \textsc{PE-Offline} & \textbf{235.23} & \textbf{628.8} & \textbf{805.47} & \textbf{1,332.00} & 4,233.83 & \textbf{0.12} & \textbf{0.27} & \textbf{0.47} & \textbf{0.93} & \textbf{1.43} \\ \cline{2-12}
& \textsc{TrieHI} & 273.10 & 667.13 & 878.33 & 1,526.20 & \textbf{4,004.53} & 0.43 & 0.70 & 1.03 & 2.80 & 3.43 \\\hline
\multirow{3}{*}{\textbf{Non-Recur.}} & \textsc{PE-Online} & \textbf{29.07} & \textbf{60.67} & \textbf{118.27} & \textbf{361.57} & \textbf{993.47} & \textbf{4.35} & \textbf{15.43} & \textbf{21.70} & \textbf{28.50} & \textbf{31.17} \\ \cline{2-12}
& \textsc{PE-Offline} & 1,037.07 & 3,029.43 & 3,935.93 & 5,987.97 & 8,228.40 & 163.91 & 25.00 & 34.70 & 3,548.13 & 3,594.70 \\ \cline{2-12}
& \textsc{TrieHI} & 945.17 & 2,467.57 & 3,419.33 & 5,097.20 & 7,528.43 & 112.49 & 21.17 & 1,568.23 & 1,980.40 & 2,094.00 \\\hline
\end{tabular*}
\vspace{-0.2cm}
\end{table*}

\vspace{0.12cm}
\noindent\textbf{Recursive DSQ.}
Figure~\ref{fig:recursive_query_performance} shows that \textsc{PE-Offline} and \textsc{TrieHI} provide the favorable accuracy-latency trade-off for recursive retrieval. This matches the analysis in Sections~\ref{sec:pe-offline} and~\ref{sec:trie-analysis}: \textsc{PE-Offline} resolves a recursive scope through a materialized ancestor key, while \textsc{TrieHI} reaches the target directory by path traversal and reads the node aggregate. Both avoid enumerating descendant path keys at query time. Table~\ref{tab:bitmap_latency} isolates this effect. On WIKI-Dir, where shallow directories can contain very large subtrees, \textsc{PE-Online} spends 365 ms on directory-only scope resolution, whereas \textsc{PE-Offline} and \textsc{TrieHI} require only 235--273 $\mu$s. The same pattern holds on ARXIV-Dir, although the hierarchy is shallower and the absolute expansion cost is smaller. Thus, query-time path expansion is the main scalability bottleneck for recursive DSQ.

\vspace{0.12cm}
\noindent\textbf{Non-Recursive DSQ.}
Non-recursive retrieval exhibits the opposite local trade-off. A non-recursive query asks for entries directly bound to a directory, so \textsc{PE-Online} can resolve the scope by a direct path-key lookup. \textsc{PE-Offline} and \textsc{TrieHI} compute the same scope by subtracting child-subtree aggregates, which introduces set-difference work proportional to immediate fan-out. Table~\ref{tab:bitmap_latency} reports this overhead: on WIKI-Dir, \textsc{PE-Online} needs 29 $\mu$s on average, while \textsc{PE-Offline} and \textsc{TrieHI} are near 1 ms. However, Figure~\ref{fig:non_recursive_query_performance} shows that this metadata cost remains small relative to end-to-end vector search. The result clarifies the trade-off: \textsc{TrieHI} is not designed to minimize the simplest non-recursive lookup alone; it aims to provide efficient recursive scopes while keeping non-recursive overhead bounded.

\begin{figure}
  \setlength{\abovecaptionskip}{0.2cm}
  \setlength{\belowcaptionskip}{-0.1cm}
  \centering
  \footnotesize
  \stackunder[1pt]{\includegraphics[scale=0.28]{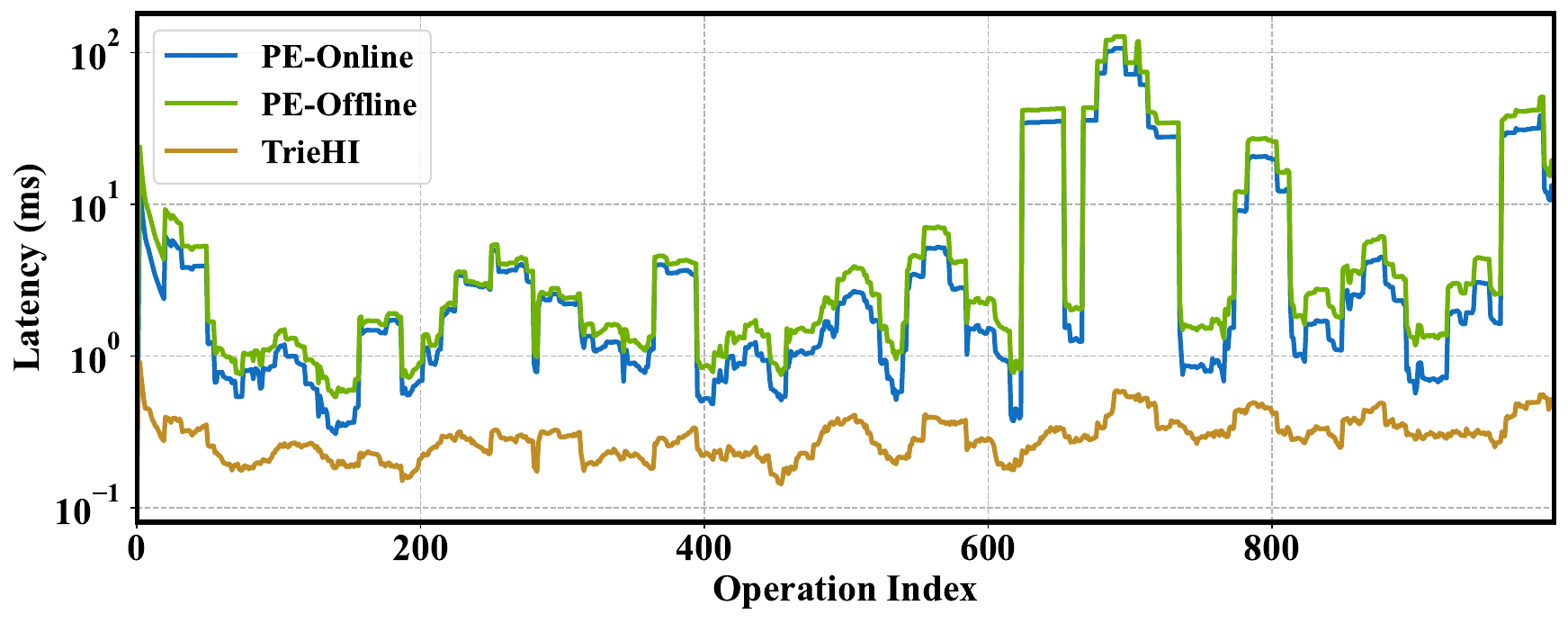}}{\hspace{0.2cm}(a) \texttt{MOVE}}
  \newline
  \stackunder[1pt]{\includegraphics[scale=0.28]{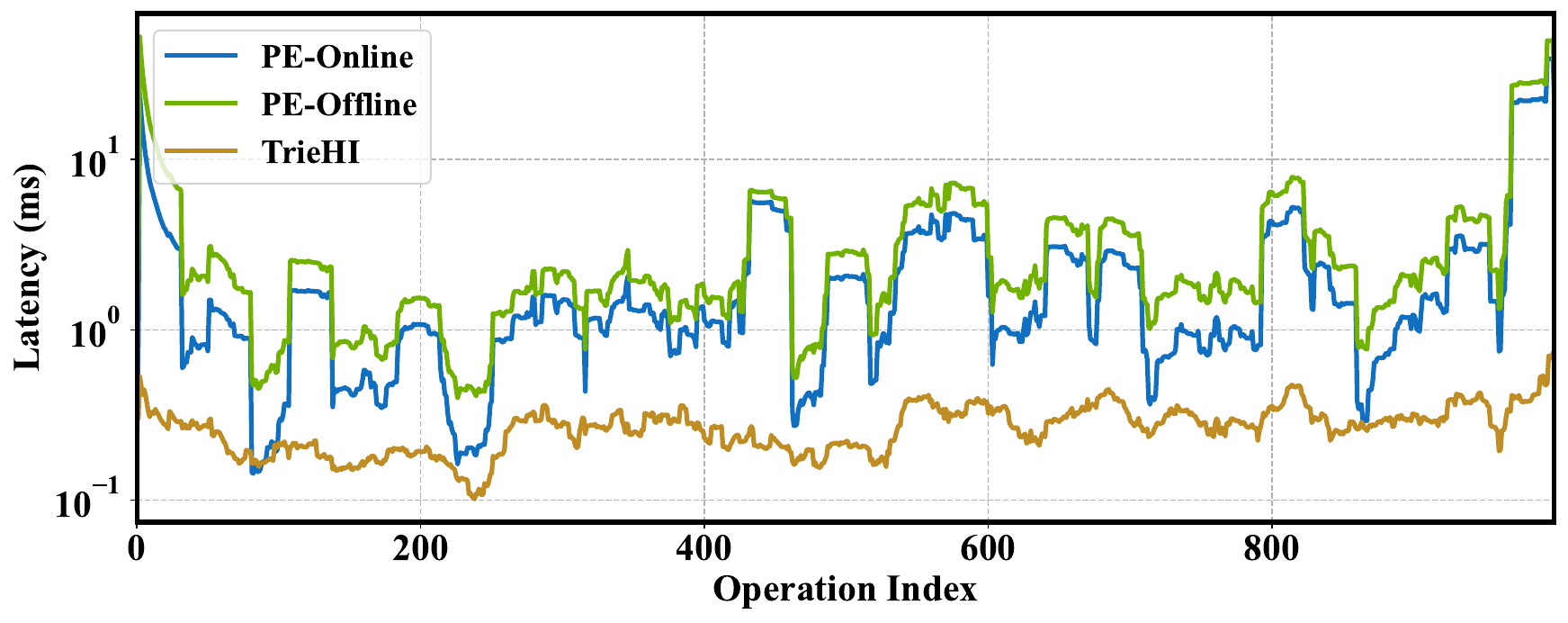}}{\hspace{0.2cm}(b) \texttt{MERGE}}
  \newline
  \vspace{-0.1cm}
  \caption{Wall-clock latency for DSM operations.}
  \label{fig:dsm_latency}
  \vspace{-0.4cm}
\end{figure}

\subsection{Directory-Semantic Maintenance}
\label{sec:maint-perf}

Figure~\ref{fig:dsm_latency} evaluates DSM operations. \textsc{TrieHI} avoids subtree-wide path-key rewriting for moves and limits merge work to the branches that must be reconciled. A \texttt{MOVE} relinks the source subtree and updates affected ancestor aggregates. Its metadata work is determined by the source and destination paths, plus bitmap updates on the moved entry set, rather than by the number of descendant directory keys. A \texttt{MERGE} may recursively visit conflicting branches, but non-conflicting child subtrees can be transferred as node units.
The expansion-based designs expose the cost of representing topology as scalar path keys. Moving or merging a directory requires updating the path-key representation of affected descendants or their materialized ancestor memberships. As the subtree grows, these updates become increasingly expensive and produce high latency variance. The DSM results therefore provide the main evidence for the native-tree design: \textsc{TrieHI} does not merely accelerate recursive scope resolution; it represents each directory as a trie node, so structural mutations can be expressed as subtree relinking, conflict-local merging, and ancestor-aggregate updates.

\begin{table*}[t!]
\centering
\caption{Index construction time and size, measured with the same baseline vector index.}
\vspace{-0.2cm}
\label{tab:indexing}
\scriptsize
\setlength{\tabcolsep}{.029\linewidth}
\renewcommand{\arraystretch}{1.16}
\begin{tabular*}{\textwidth}{@{\extracolsep{\fill}}>{\raggedright\arraybackslash}m{0.14\textwidth}|cccc|cccc@{}}
\hline
\textbf{Dataset}$\rightarrow$ & \multicolumn{4}{c|}{\textbf{WIKI-Dir}} & \multicolumn{4}{c}{\textbf{ARXIV-Dir}} \\ \hline
\textbf{Metric}$\rightarrow$ & \multicolumn{2}{c}{\textbf{Indexing Time (s)}} & \multicolumn{2}{c|}{\textbf{Index Size (MB)}} & \multicolumn{2}{c}{\textbf{Indexing Time (s)}} & \multicolumn{2}{c}{\textbf{Index Size (MB)}} \\ \hline
\textbf{Vector Index}$\rightarrow$ & \textbf{PG} & \textbf{IVF} & \textbf{PG} & \textbf{IVF} & \textbf{PG} & \textbf{IVF} & \textbf{PG} & \textbf{IVF} \\ \hline
Baseline (Vec.) & 1491.27 & 442.32 & 7,925 & 7,744 & 720.13 & 199.94 & 11,227 & 10,964 \\ \hline
\textsc{PE-Online} (Vec.+Dir.) & 1502.49 & 453.53 & 8,143 & 7,962 & 729.07 & 208.89 & 11,239 & 10,976 \\ \hline
\textsc{PE-Offline} (Vec.+Dir.) & 1,515.25 & 466.29 & 8,213 & 8,032 & 733.62 & 213.43 & 11,241 & 10,978 \\ \hline
\textsc{TrieHI} (Vec.+Dir.) & 1,502.21 & 453.25 & 8,326 & 8,145 & 729.60 & 209.41 & 11,519 & 11,256 \\ \hline
\end{tabular*}
\vspace{-0.2cm}
\end{table*}

\subsection{Indexing and Storage Overhead}
\label{sec:index-perf}
Table~\ref{tab:indexing} reports resource overhead. All directory-semantic methods add only slight construction-time overhead relative to the vector-index baseline. On WIKI-Dir with PG, the baseline takes 1491.27s, while the three directory-aware variants take 1502.21--1515.25s, an increase of less than 1.7\%.

The storage results show the expected differences among the directory representations. \textsc{PE-Online} has the smallest footprint because it stores only immediate path memberships. \textsc{PE-Offline} spends additional space on ancestor materialization; this cost is more visible on WIKI-Dir, whose paths are deeper. \textsc{TrieHI} stores trie topology and per-node inclusive entry-ID sets, giving it the largest storage overhead in both datasets. This overhead is the price paid for reusable directory scopes and tree-local maintenance. The results therefore do not identify a single universally dominant representation; rather, they show that \textsc{TrieHI} trades additional metadata space for robust recursive DSQ and efficient DSM.

\begin{figure}[t]
\centering
\includegraphics[width=0.9\linewidth]{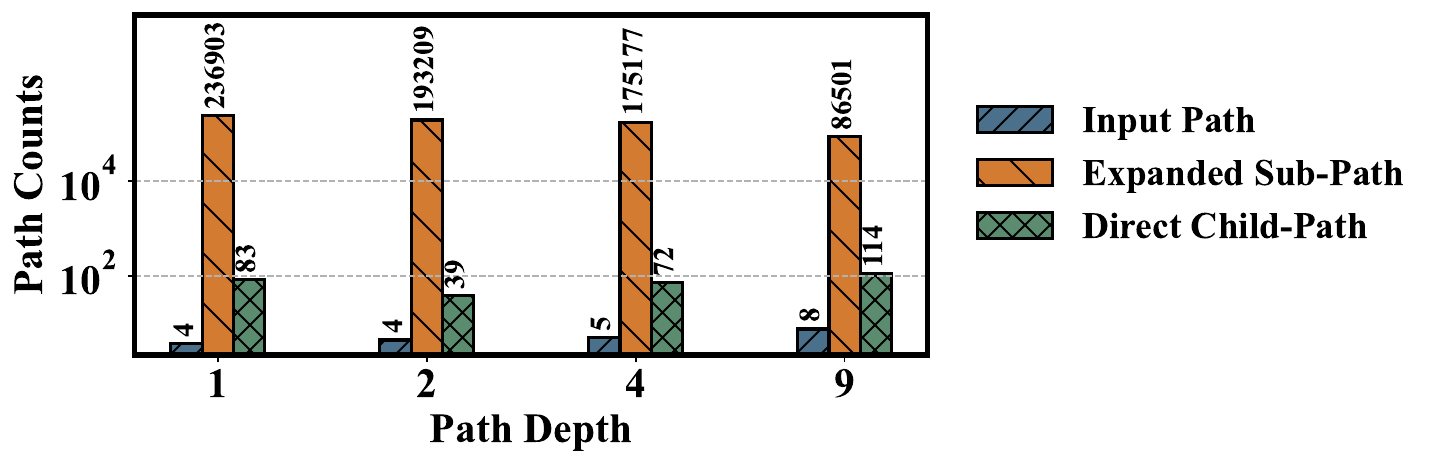}
\vspace{-0.2cm}
\caption{WIKI-Dir structural complexity by path depth. ``Input Path'' is the number of average query anchor directories at that depth; ``Expanded Sub-Path'' ($\boldsymbol{m}$) is the number of descendant directory keys expanded from those anchors; ``Direct Child-Path'' ($\boldsymbol{c}$) is the number of immediate child directories used for non-recursive set difference.}
\label{fig:path_counts}
\vspace{-0.4cm}
\end{figure}

\begin{figure}
  \setlength{\abovecaptionskip}{0.2cm}
  \setlength{\belowcaptionskip}{-0.1cm}
  \centering
  \footnotesize
  \hspace{0.4cm}
  \stackunder[1pt]{\includegraphics[scale=0.34]{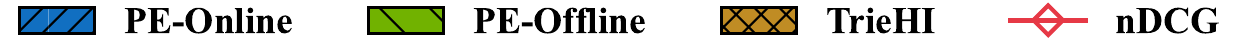}}{}
  \newline
  \stackunder[1pt]{\includegraphics[scale=0.28]{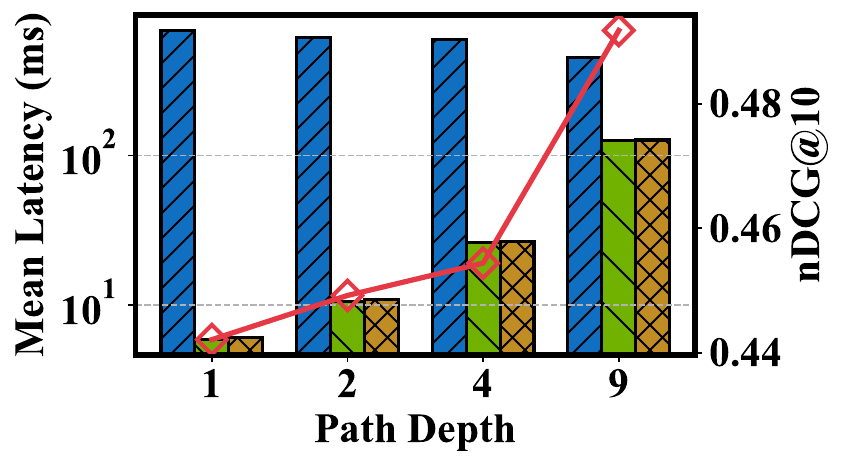}}{\hspace{0.2cm}(a) PG}
  \hspace{0.25cm}
  \stackunder[1pt]{\includegraphics[scale=0.28]{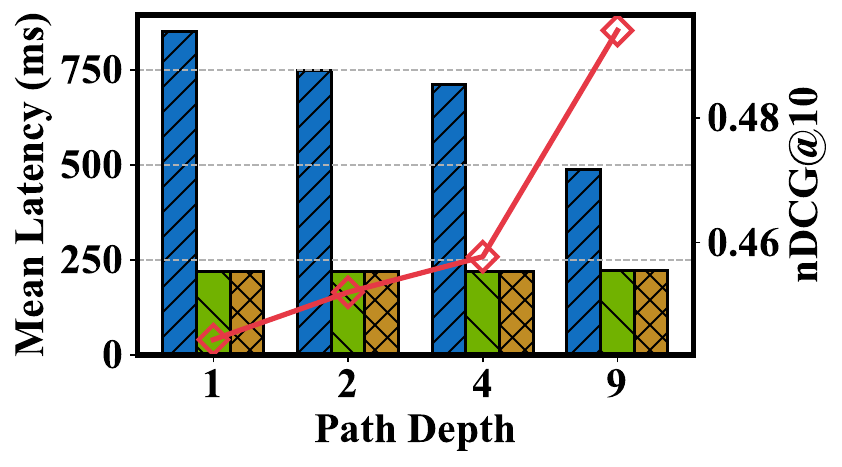}}{\hspace{0.2cm}(b) IVF}
  \newline
  \vspace{-0.2cm}
  \caption{Effect of query path depth on recursive DSQ performance on WIKI-Dir.}
  \label{fig:recur_query_perf_depth}
  \vspace{-0.2cm}
\end{figure}

\begin{figure}
  \setlength{\abovecaptionskip}{0.2cm}
  \setlength{\belowcaptionskip}{-0.1cm}
  \centering
  \footnotesize
  \stackunder[1pt]{\includegraphics[scale=0.3]{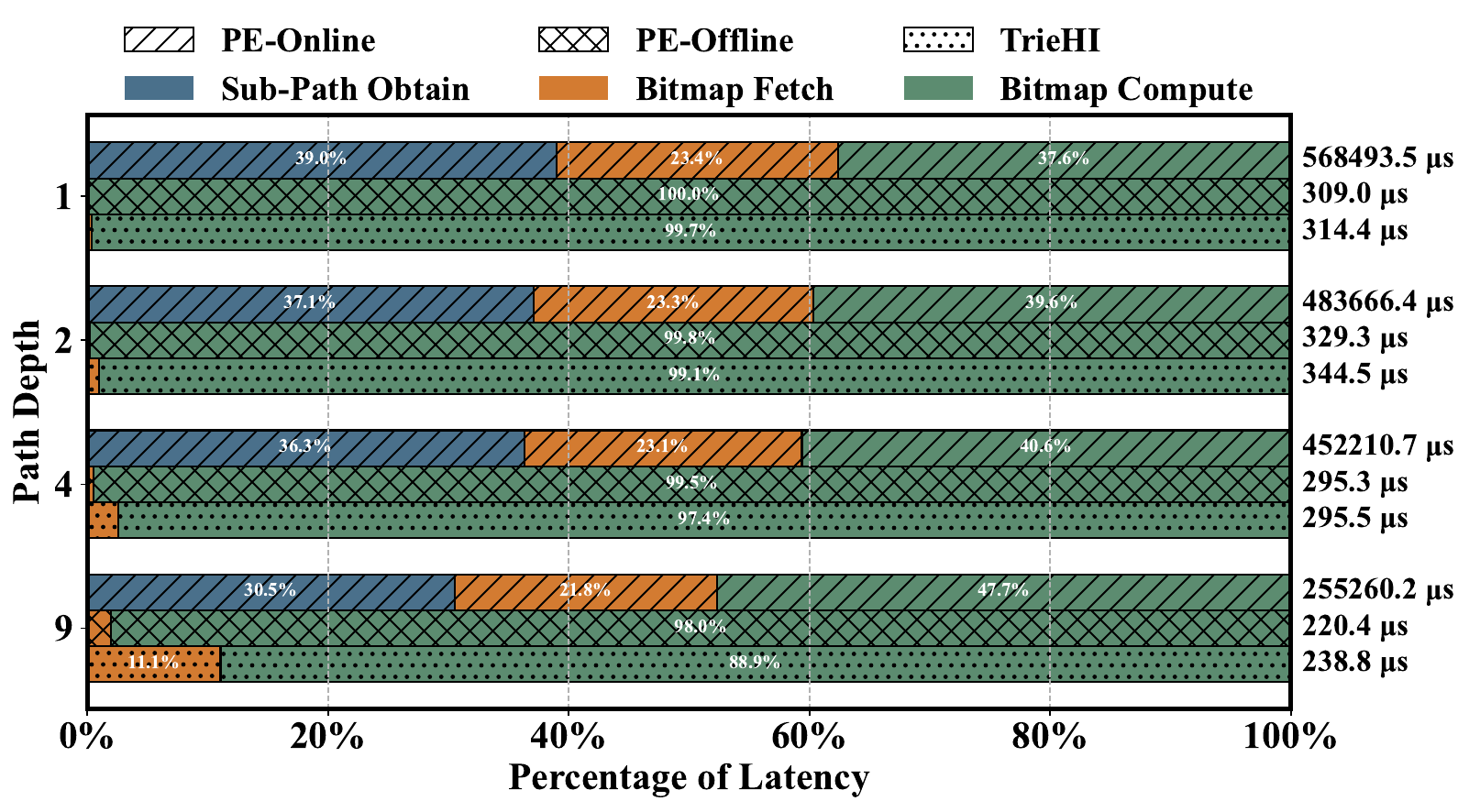}}{\hspace{0.2cm}(a) Recursive Query}
  \newline
  \stackunder[1pt]{\includegraphics[scale=0.3]{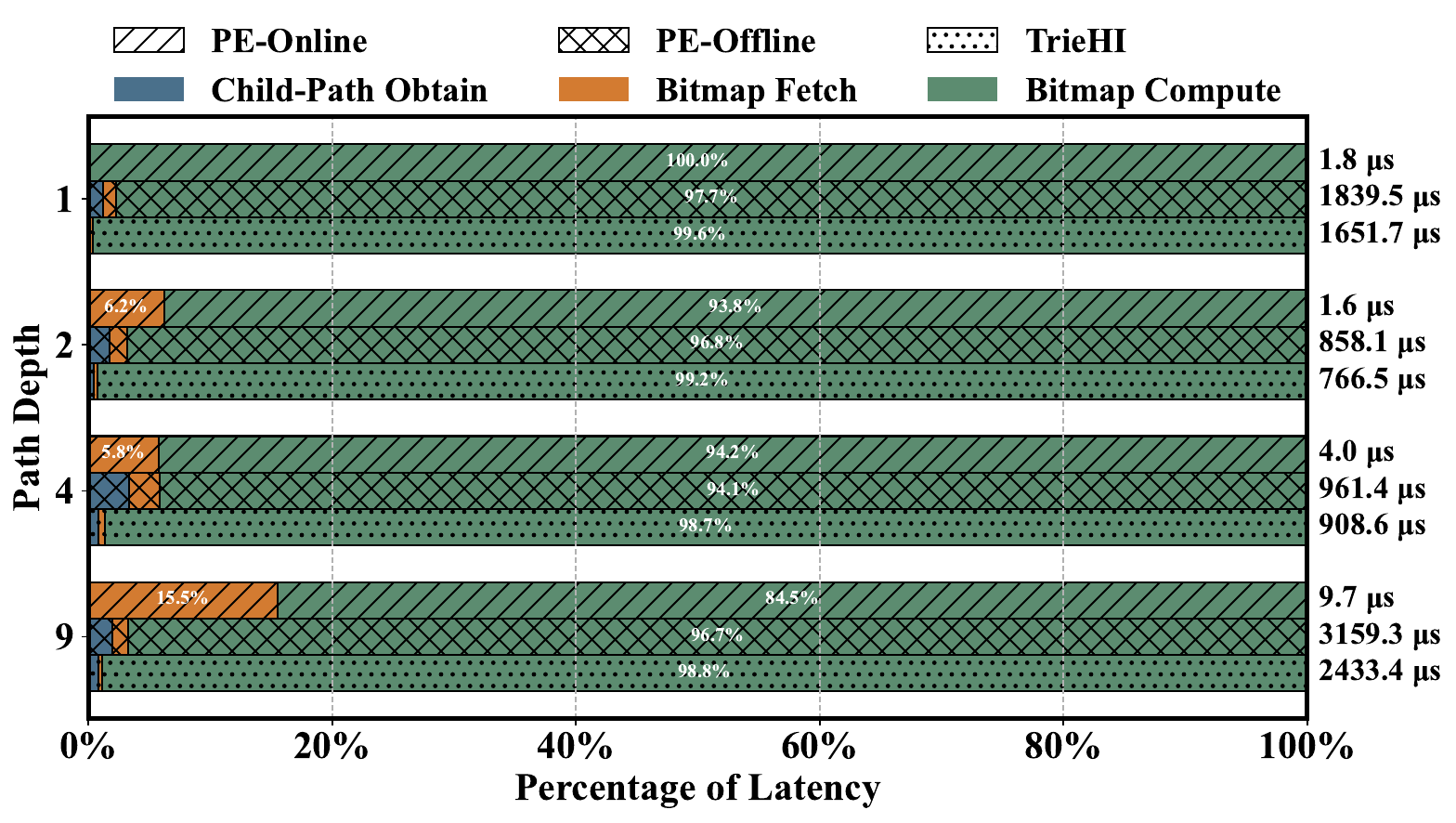}}{\hspace{0.2cm}(b) Non-Recursive Query}
  \newline
  \caption{Directory-only latency decomposition on WIKI-Dir by query path depth.}
  \label{fig:latency_breakdown}
  \vspace{-0.2cm}
\end{figure}

\subsection{Depth Sensitivity and Latency Breakdown}
\label{sec:detail-eval}

The depth analysis connects the empirical results to the cost variables in Sections~\ref{sec:expansion} and~\ref{sec:trie}. Figure~\ref{fig:path_counts} shows that shallow WIKI-Dir paths contain very large descendant subtrees, while immediate fan-out remains comparatively bounded. This structure is precisely where query-time expansion is vulnerable.

Figure~\ref{fig:recur_query_perf_depth} shows that retrieval quality generally improves as the query path becomes more specific, confirming that directory scope carries useful retrieval context. The latency trends separate the designs. \textsc{PE-Online} is slowest at shallow depths because it must enumerate many descendant paths. As depth increases and the descendant scope narrows, its overhead falls. \textsc{PE-Offline} and \textsc{TrieHI} avoid this query-time expansion. After scope resolution, the remaining latency is determined by how the ANN executor searches within the resolved candidate set. For PG, their latency increases with depth because deeper directory constraints are more selective: the candidate mask reduces the density of valid nodes in PG and weakens graph connectivity, so the search may perform more traversal work to collect valid results. For IVF, \textsc{PE-Offline} and \textsc{TrieHI} show a flatter latency profile: partition probing dominates the end-to-end cost, and directory-scope resolution adds little depth-dependent overhead. Thus, PG and IVF exhibit different latency curves. Across both executors, the key requirement is to avoid descendant-path enumeration; executor-specific latency then depends on how PG or IVF consumes the resolved candidate set.

Figure~\ref{fig:latency_breakdown} decomposes directory-only latency. For recursive DSQ, \textsc{PE-Online} follows the expected descendant-enumeration profile: shallow anchors expand to many descendant keys, so ``Sub-Path Obtain'' and ``Bitmap Fetch'' dominate, and latency decreases as the descendant scope narrows with depth. \textsc{PE-Offline} and \textsc{TrieHI} avoid this enumeration; recursive scope resolution becomes an ancestor-materialized lookup or a trie traversal plus aggregate-set access, keeping directory-only latency in the hundreds of microseconds. For non-recursive DSQ, the pattern reverses. \textsc{PE-Online} uses a direct exact-parent lookup, whereas \textsc{PE-Offline} and \textsc{TrieHI} exclude immediate child subtrees through union and set difference, shown as ``Bitmap Compute''. Because immediate fan-out is much smaller and more stable than the full descendant scope, this extra work remains predictable.

\subsection{End-to-End Evaluation in \textsf{OpenViking}}
\label{sec:openviking-case}
We now evaluate the application effect of \textsc{TrieHI} in \textsf{OpenViking} on agent-context and knowledge-base QA workloads.

\vspace{0.10cm}
\noindent\textbf{Evaluation Setup.}
LOCOMO~\cite{LOCOMO} evaluates long-conversation user-memory QA for OpenClaw~\cite{openclaw}, Hermes~\cite{hermes}, and Claude Code~\cite{claudecode}, comparing each native-memory baseline with its \textsf{OpenViking} variant. HotpotQA~\cite{hotpotqa} evaluates multi-hop knowledge-base QA against direct retrieval baselines, including Naive RAG, HippoRAG 2~\cite{gutierrez2025rag,gutierrez2024hipporag}, and LightRAG~\cite{LightRAG}; \textsf{OpenViking} is measured with top-5 and top-20 retrieval budgets. The benchmark uses an LLM-as-judge protocol with \texttt{doubao-seed-2-0-pro-260215}; detailed configurations are available in the \textsf{OpenViking} benchmark repository\footnote{\url{https://github.com/volcengine/OpenViking/tree/main/benchmark}}. Accuracy denotes judged answer correctness. Avg. Query Time and Latency measure retrieval or memory-access time per question. Total Input Tokens sums input tokens over LOCOMO queries, while Tokens/QA is the average input-token cost per HotpotQA question.

\vspace{0.10cm}
\noindent\textbf{User Memory.}
Table~\ref{tab:openviking_locomo} evaluates long-conversation user memory on the LOCOMO dataset. Across OpenClaw, Hermes, and Claude Code, integrating \textsf{OpenViking} improves QA accuracy and reduces both average query time and total input-token usage. \textsc{TrieHI} supplies the scope-resolution step: a query first resolves the relevant user or session scope, and vector ranking then operates over that scoped candidate set. Together with L0/L1/L2 context levels, this avoids repeatedly injecting broad conversation history into the prompt while still allowing recursive access to detailed evidence when needed.

\begin{table}[t]
\centering
\caption{User-memory QA performance on LOCOMO.}
\vspace{-0.2cm}
\label{tab:openviking_locomo}
\begingroup
\scriptsize
\setlength{\tabcolsep}{3pt}
\renewcommand{\arraystretch}{1.16}
\begin{tabular*}{\columnwidth}{@{\extracolsep{\fill}}>{\raggedright\arraybackslash}m{0.36\columnwidth}|>{\raggedleft\arraybackslash}m{0.13\columnwidth}|>{\raggedleft\arraybackslash}m{0.16\columnwidth}|>{\raggedleft\arraybackslash}m{0.20\columnwidth}@{}}
\hline
\textbf{Integration} & \textbf{Accuracy} & \textbf{\makecell{Avg. Query\\Time (s)}} & \textbf{\makecell{Total Input\\Tokens}} \\ \hline
OpenClaw native memory & 24.20\% & 95.14 & 392,559,404 \\
OpenClaw + \textsf{OpenViking} & \textbf{82.08\%} & \textbf{38.8} & \textbf{37,423,456} \\ \hline
Hermes native memory & 33.38\% & 82.4 & 79,228,398 \\
Hermes + \textsf{OpenViking} & \textbf{82.86\%} & \textbf{27.9} & \textbf{52,026,755} \\ \hline
Claude Code auto-memory & 57.21\% & 49.1 & 353,306,422 \\
Claude Code + \textsf{OpenViking} & \textbf{80.32\%} & \textbf{20.4} & \textbf{129,968,899} \\ \hline
\end{tabular*}
\endgroup
% \vspace{-0.2cm}
\end{table}

\vspace{0.10cm}
\noindent\textbf{Multi-Hop Knowledge-Base QA.}
HotpotQA requires evidence from multiple facts, making it a natural test for directory-scoped retrieval. In Table~\ref{tab:openviking_hotpot_retriever}, \textsf{OpenViking} top-20 obtains 91.00\% accuracy with 0.23s retrieval latency per query. Compared with LightRAG, it achieves higher accuracy, uses fewer tokens per query, and avoids the high retrieval latency observed in graph-based pipelines. This behavior matches the DSQ design: \textsc{TrieHI} resolves a coherent directory scope before vector ranking, so increasing the top-$k$ budget mainly adds related in-scope evidence rather than unrelated corpus-wide passages. The top-5/top-20 comparison shows this coverage effect while retrieval latency remains low.

\begin{table}[t]
\centering
\caption{Multi-hop RAG performance on HotpotQA.}
\vspace{-0.2cm}
\label{tab:openviking_hotpot_retriever}
\begingroup
\scriptsize
\setlength{\tabcolsep}{5pt}
\renewcommand{\arraystretch}{1.16}
\begin{tabular*}{\columnwidth}{@{\extracolsep{\fill}}>{\raggedright\arraybackslash}m{0.30\columnwidth}|>{\raggedleft\arraybackslash}m{0.13\columnwidth}|>{\raggedleft\arraybackslash}m{0.16\columnwidth}|>{\raggedleft\arraybackslash}m{0.24\columnwidth}@{}}
\hline
\textbf{Method} & \textbf{Accuracy} & \textbf{Tokens / QA} & \textbf{Latency (s) / QA} \\ \hline
Naive RAG & 62.50\% & 1,290 & \textbf{0.11} \\
HippoRAG 2 & 61.00\% & \textbf{726} & 20 \\
LightRAG & 89.00\% & 28,443 & 75 \\ \hline
\textsf{OpenViking} (top-5) & 72.75\% & 3,154 & 0.22 \\
\textsf{OpenViking} (top-20) & \textbf{91.00\%} & 12,533 & 0.23 \\ \hline
\end{tabular*}
\endgroup
\vspace{-0.2cm}
\end{table}

\section{Related Work}
\label{sec:related}
This work is related to ANN indexing, filtered vector search, hybrid retrieval, and hierarchical knowledge organization.

\vspace{0.2cm}
\noindent\textbf{ANN Indexing and Vector Databases.}
Approximate nearest neighbor (ANN) search has a long line of work on reducing vector-search latency, memory footprint, and construction cost. Representative techniques include tree-based partitioning~\cite{ram2019revisiting,DasguptaF08,WangWJLZZH14}, hashing~\cite{GionisIM99,LiZSWT020,zheng2020pm,huang2015query}, quantization~\cite{zhang2014composite,gao2024rabitq,jegou2010product,ge2013optimized}, and graph-based indexes such as HNSW~\cite{malkov2018efficient,wang2021comprehensive,fu2017fast,chen2025maximum,peng2023efficient,wang2025boosting,fu2025vista,li2025fast}. Recent systems further optimize memory access~\cite{WangWKGZZ25,VSAG,GouGXL25}, distance computation~\cite{gao2023high,YangLJZWSJW25,DengCZWZZ24,li2020improving}, construction~\cite{YangXLYGWPC25,abs-2508-08744}, disk-resident search~\cite{YueZXXZDGZJ25,wang2024starling,jayaram2019diskann}, and dynamic updates~\cite{LiuZYRZJ25,MaZZCGG24,abs-2502-13826}.
These techniques improve how vectors are ranked once a candidate space is defined. Our work addresses a different layer: how a hierarchical directory constraint is resolved into the candidate entry set supplied to the ANN executor. Our strategies are independent of the ANN layout and can work with both graph-based and partition-based vector indexes, as evaluated in Section~\ref{sec:experiment}.

\vspace{0.2cm}
\noindent\textbf{Filtered Vector Search.}
Vector databases commonly combine ANN search with scalar predicates. Decoupled systems and optimizers~\cite{ye2025compass,NaviX,ACORN,ADBV,2021milvus,HS-Apple,SIEVE} separate predicate evaluation from vector indexing and choose execution strategies according to predicate selectivity~\cite{abs-2508-16263,abs-2507-21989,abs-2505-06501,abs-2509-07789}. Other work builds coupled indexes for specific predicate forms, such as tags~\cite{LuoQZD25,GollapudiKSKBRL23,ACORN} and ranges~\cite{LiangZYCSC24,WangHTZZ25,abs-2508-18617,ZuoQZLD24}, or for broader hybrid filtering settings~\cite{CaiSCZ24,HQANN,NHQ,abs-2509-19767}. Directory constraints can be approximated with scalar path filters, but existing vector databases typically treat paths as flat attributes rather than hierarchy-aware constraints. Exact-parent predicates can be evaluated by equality filtering, while recursive scopes and namespace mutations must be encoded through additional metadata expansion or application-side coordination. Sections~\ref{sec:expansion} and~\ref{sec:trie} formalize this design space inside the vector-database layer: \textsc{PE-Online} and \textsc{PE-Offline} study expansion-based support for DSQ and DSM, while \textsc{TrieHI} provides a native prefix-tree representation for reusable directory scopes and structural maintenance.

\vspace{0.2cm}
\noindent\textbf{Hybrid Retrieval.}
Hybrid retrieval combines dense semantic search with sparse lexical retrieval to improve ranking quality in retrieval pipeline~\cite{BM25,abs-2508-01405,BruchNIL24a,LuanETC21,YangCHQY24,Blend-RAG,li2025all}. Existing work studies rank fusion, including Reciprocal Rank Fusion and convex combinations~\cite{ChenZLBN22,HS-FusionFunctions}, as well as joint dense-sparse indexing~\cite{HS-HybridIndex,OneSparse,li2025all}. These methods decide how to rank or fuse evidence after retrieval signals are produced.
Directory semantics solve a different problem: they define which entries are eligible for retrieval under a hierarchical scope. A DSQ can first resolve the valid candidate set for a directory subtree, after which dense, sparse, or hybrid retrieval can rank within that set. This makes our approach complementary to relevance-oriented hybrid retrieval rather than a replacement for it.

\vspace{0.2cm}
\noindent\textbf{Hierarchical Knowledge Organization.}
Directories and hierarchical namespaces are widely used to organize code repositories, enterprise knowledge bases, personal files, and agent-managed context~\cite{lynch2025directories,JacobsonKSS98,LutzRBD14,CodeSearch,Enterprise-KB}. RAG and agent systems increasingly rely on such structure to manage external context~\cite{ragflow,openclaw}, but the hierarchy is typically maintained as application-side metadata. Recursive retrieval then becomes an external preprocessing step, and namespace updates are coordinated outside the vector database.
Our work brings this organization model into the vector-database layer. DSQ captures directory-scoped retrieval, and DSM captures topology-preserving updates. The \textsf{OpenViking} integration in Section~\ref{sec:openviking-integration} illustrates how the same abstraction supports filesystem-style context organization and directory-recursive retrieval in an agent context database.
\section{Conclusion}
\label{sec:conclusion}
This paper studies directory semantics as a first-class capability for vector databases. We formalized Directory-Semantic Query (DSQ) and Directory-Semantic Maintenance (DSM), analyzed two expansion-based implementations, and introduced \textsc{TrieHI}, a native prefix-tree metadata index that preserves directory topology above the ANN executor. The evaluation on WIKI-Dir and ARXIV-Dir shows the core trade-off: path expansion can reuse scalar metadata interfaces but incurs high expansion cost and subtree-wide maintenance amplification, whereas \textsc{TrieHI} resolves recursive scopes through directory nodes and supports structural updates through tree-local operations. The \textsf{OpenViking} integration further shows that this abstraction can support filesystem-style context organization and directory-recursive retrieval in agent-memory and knowledge-base QA workloads. Future work includes extending DSQ to richer path predicates and cost-aware planning for mixed semantic and hierarchical constraints.

\bibliographystyle{IEEEtran}
\balance
\bibliography{main.bib}

\begin{thebibliography}{100}
\providecommand{\url}[1]{#1}
\csname url@samestyle\endcsname
\providecommand{\newblock}{\relax}
\providecommand{\bibinfo}[2]{#2}
\providecommand{\BIBentrySTDinterwordspacing}{\spaceskip=0pt\relax}
\providecommand{\BIBentryALTinterwordstretchfactor}{4}
\providecommand{\BIBentryALTinterwordspacing}{\spaceskip=\fontdimen2\font plus
\BIBentryALTinterwordstretchfactor\fontdimen3\font minus \fontdimen4\font\relax}
\providecommand{\BIBforeignlanguage}[2]{{%
\expandafter\ifx\csname l@#1\endcsname\relax
\typeout{** WARNING: IEEEtran.bst: No hyphenation pattern has been}%
\typeout{** loaded for the language `#1'. Using the pattern for}%
\typeout{** the default language instead.}%
\else
\language=\csname l@#1\endcsname
\fi
#2}}
\providecommand{\BIBdecl}{\relax}
\BIBdecl

\bibitem{Survey-VDB-VLDBJ24}
J.~J. Pan, J.~Wang, and G.~Li, ``Survey of vector database management systems,'' \emph{The VLDB Journal}, vol.~33, no.~5, pp. 1591--1615, 2024.

\bibitem{SingleStore-V}
C.~Chen, C.~Jin, Y.~Zhang, S.~Podolsky, C.~Wu, S.~Wang, E.~Hanson, Z.~Sun, R.~Walzer, and J.~Wang, ``Singlestore-v: An integrated vector database system in singlestore,'' \emph{Proc. {VLDB} Endow.}, vol.~17, no.~12, pp. 3772--3785, 2024.

\bibitem{AlayaDB}
Y.~Deng, Z.~You, L.~Xiang, Q.~Li, P.~Yuan, Z.~Hong, Y.~Zheng, W.~Li, R.~Li, H.~Liu, K.~Mouratidis, M.~L. Yiu, H.~Li, Q.~Shen, R.~Mao, and B.~Tang, ``Alayadb: The data foundation for efficient and effective long-context {LLM} inference,'' in \emph{Companion of the International Conference on Management of Data (SIGMOD)}, 2025, pp. 364--377.

\bibitem{HAKES}
G.~Hu, S.~Cai, T.~T.~A. Dinh, Z.~Xie, C.~Yue, G.~Chen, and B.~C. Ooi, ``{HAKES:} scalable vector database for embedding search service,'' \emph{Proc. {VLDB} Endow.}, vol.~18, no.~9, pp. 3049--3062, 2025.

\bibitem{MicroNN}
J.~Pound, F.~Chabert, A.~Bhushan, A.~Goswami, A.~Pacaci, and S.~R. Chowdhury, ``Micronn: An on-device disk-resident updatable vector database,'' in \emph{Companion of the International Conference on Management of Data (SIGMOD)}, 2025, pp. 608--621.

\bibitem{VSAG}
X.~Zhong, H.~Li, J.~Jin, M.~Yang, D.~Chu, X.~Wang, Z.~Shen, W.~Jia, G.~Gu, Y.~Xie, X.~Lin, H.~T. Shen, J.~Song, and P.~Cheng, ``{VSAG:} an optimized search framework for graph-based approximate nearest neighbor search,'' \emph{Proc. {VLDB} Endow.}, vol.~18, no.~12, pp. 5017--5030, 2025.

\bibitem{Enterprise-KB}
F.~Jiang, C.~Qin, K.~Yao, C.~Fang, F.~Zhuang, H.~Zhu, and H.~Xiong, ``Enhancing question answering for enterprise knowledge bases using large language models,'' in \emph{Database Systems for Advanced Applications - International Conference (DASFAA)}, vol. 14853, 2024, pp. 273--290.

\bibitem{Wiki-WebSearch}
M.~Liu, S.~Zhong, Q.~Yang, Y.~Han, X.~Liu, and Y.~Ma, ``Webanns: Fast and efficient approximate nearest neighbor search in web browsers,'' in \emph{Proceedings of the International {ACM} {SIGIR} Conference on Research and Development in Information Retrieval}, 2025, pp. 2483--2492.

\bibitem{arXiv-VDB}
S.~S. Monir, I.~Lau, S.~Yang, and D.~Zhao, ``Vectorsearch: Enhancing document retrieval with semantic embeddings and optimized search,'' \emph{arXiv:2409.17383}, 2024.

\bibitem{CodeSearch}
J.~Chen, X.~Hu, Z.~Li, C.~Gao, X.~Xia, and D.~Lo, ``Code search is all you need? improving code suggestions with code search,'' in \emph{Proceedings of the {IEEE/ACM} International Conference on Software Engineering (ICSE)}, 2024, pp. 73:1--73:13.

\bibitem{Chat2Data}
X.~Zhao, X.~Zhou, and G.~Li, ``Chat2data: An interactive data analysis system with rag, vector databases and llms,'' \emph{Proc. {VLDB} Endow.}, vol.~17, no.~12, pp. 4481--4484, 2024.

\bibitem{LLM-DB-Survey}
Z.~Jing, Y.~Su, Y.~Han, B.~Yuan, H.~Xu, C.~Liu, K.~Chen, and M.~Zhang, ``When large language models meet vector databases: {A} survey,'' \emph{arXiv:2402.01763}, 2024.

\bibitem{DB-Debug-RAG}
S.~Chen, J.~Fan, B.~Wu, N.~Tang, C.~Deng, P.~Wang, Y.~Li, J.~Tan, F.~Li, J.~Zhou, and X.~Du, ``Automatic database configuration debugging using retrieval-augmented language models,'' \emph{Proc. {ACM} Manag. Data}, vol.~3, no.~1, pp. 13:1--13:27, 2025.

\bibitem{MQA}
M.~Wang, H.~Wu, X.~Ke, Y.~Gao, X.~Xu, and L.~Chen, ``An interactive multi-modal query answering system with retrieval-augmented large language models,'' \emph{Proc. {VLDB} Endow.}, vol.~17, no.~12, pp. 4333--4336, 2024.

\bibitem{claudecode}
``Claude code,'' \url{https://github.com/anthropics/claude-code}, 2026, [Online; accessed 05-May-2026].

\bibitem{openclaw}
``Openclaw — personal ai assistant,'' \url{https://github.com/openclaw/openclaw}, 2026, [Online; accessed 01-February-2026].

\bibitem{hermes}
``Hermes agent,'' \url{https://github.com/nousresearch/hermes-agent}, 2026, [Online; accessed 05-May-2026].

\bibitem{HS-Milvus}
``A review of hybrid search in milvus,'' \url{https://zilliz.com/blog/a-review-of-hybrid-search-in-milvus}, 2025, [Online; accessed 20-September-2025].

\bibitem{HS-turbopuffer}
``Hybrid search,'' \url{https://turbopuffer.com/docs/hybrid}, 2025, [Online; accessed 20-September-2025].

\bibitem{HS-pinecone}
``Getting started with hybrid search,'' \url{https://www.pinecone.io/learn/hybrid-search-intro/}, 2023, [Online; accessed 06-February-2025].

\bibitem{HS-weaviate}
``Hybrid search explained,'' \url{https://weaviate.io/blog/hybrid-search-explained}, 2025, [Online; accessed 06-February-2025].

\bibitem{BM25}
S.~E. Robertson, S.~Walker, S.~Jones, M.~Hancock{-}Beaulieu, and M.~Gatford, ``Okapi at {TREC-3},'' in \emph{Proceedings of The Third Text REtrieval Conference (TREC)}, vol. 500-225, 1994, pp. 109--126.

\bibitem{HS-FusionFunctions}
S.~Bruch, S.~Gai, and A.~Ingber, ``An analysis of fusion functions for hybrid retrieval,'' \emph{{ACM} Trans. Inf. Syst.}, vol.~42, no.~1, pp. 20:1--20:35, 2024.

\bibitem{HS-HybridIndex}
H.~Zhang, J.~Liu, Z.~Zhu, S.~Zeng, M.~Sheng, T.~Yang, G.~Dai, and Y.~Wang, ``Efficient and effective retrieval of dense-sparse hybrid vectors using graph-based approximate nearest neighbor search,'' \emph{arXiv:2410.20381}, 2024.

\bibitem{OneSparse}
Y.~Chen, R.~Zheng, Q.~Chen, S.~Xu, Q.~Zhang, X.~Wu, W.~Han, H.~Yuan, M.~Li, Y.~Wang, J.~Li, F.~Yang, H.~Sun, W.~Deng, F.~Sun, Q.~Zhang, and M.~Yang, ``Onesparse: {A} unified system for multi-index vector search,'' in \emph{Companion Proceedings of the {ACM} on Web Conference (WWW)}, 2024, pp. 393--402.

\bibitem{VBase}
Q.~Zhang, S.~Xu, Q.~Chen, G.~Sui, J.~Xie, Z.~Cai, Y.~Chen, Y.~He, Y.~Yang, F.~Yang, M.~Yang, and L.~Zhou, ``{VBASE:} unifying online vector similarity search and relational queries via relaxed monotonicity,'' in \emph{{USENIX} Symposium on Operating Systems Design and Implementation (OSDI)}, 2023, pp. 377--395.

\bibitem{HS-Apple}
J.~Mohoney, A.~Pacaci, S.~R. Chowdhury, A.~Mousavi, I.~F. Ilyas, U.~F. Minhas, J.~Pound, and T.~Rekatsinas, ``High-throughput vector similarity search in knowledge graphs,'' \emph{Proc. {ACM} Manag. Data}, vol.~1, no.~2, pp. 197:1--197:25, 2023.

\bibitem{ADBV}
C.~Wei, B.~Wu, S.~Wang, R.~Lou, C.~Zhan, F.~Li, and Y.~Cai, ``Analyticdb-v: {A} hybrid analytical engine towards query fusion for structured and unstructured data,'' \emph{Proc. {VLDB} Endow.}, vol.~13, no.~12, pp. 3152--3165, 2020.

\bibitem{2021milvus}
J.~Wang, X.~Yi, R.~Guo, H.~Jin, P.~Xu, S.~Li, X.~Wang, X.~Guo, C.~Li, X.~Xu \emph{et~al.}, ``Milvus: A purpose-built vector data management system,'' in \emph{Proceedings of the International Conference on Management of Data (SIGMOD)}, 2021, pp. 2614--2627.

\bibitem{opensearch}
``Vector search,'' \url{https://opensearch.org/platform/vector-search/}, 2025, [Online; accessed 11-September-2025].

\bibitem{byteplus-intro}
``Intelligent vector infrastructure,'' \url{https://go.byteplus.com/Vikingdbvectordatabase}, 2025, [Online; accessed 10-November-2025].

\bibitem{byteplus}
``Vikingdb vector database,'' \url{https://www.byteplus.com/en/product/vectordatabase}, 2025, [Online; accessed 10-November-2025].

\bibitem{lynch2025directories}
O.~Lynch and M.~Lohmayer, ``Directories: A convenient and well-behaved formalism for hierarchical organization in categorical systems theory,'' \emph{arXiv:2504.19389}, 2025.

\bibitem{JacobsonKSS98}
G.~Jacobson, B.~Krishnamurthy, D.~Srivastava, and D.~Suciu, ``Focusing search in hierarchical structures with directory sets,'' in \emph{Proceedings of the {ACM} {CIKM} International Conference on Information and Knowledge Management}, 1998, pp. 1--9.

\bibitem{LutzRBD14}
R.~Lutz, D.~Rausch, F.~Beck, and S.~Diehl, ``Get your directories right: From hierarchy visualization to hierarchy manipulation,'' in \emph{{IEEE} Symposium on Visual Languages and Human-Centric Computing (VL/HCC)}, 2014, pp. 25--32.

\bibitem{DBPedia}
F.~Hasibi, F.~Nikolaev, C.~Xiong, K.~Balog, S.~E. Bratsberg, A.~Kotov, and J.~Callan, ``Dbpedia-entity v2: A test collection for entity search,'' in \emph{Proceedings of the International ACM SIGIR Conference on Research and Development in Information Retrieval}, 2017, pp. 1265--1268.

\bibitem{BGE-M3}
J.~Chen, S.~Xiao, P.~Zhang, K.~Luo, D.~Lian, and Z.~Liu, ``{BGE} m3-embedding: Multi-lingual, multi-functionality, multi-granularity text embeddings through self-knowledge distillation,'' \emph{arXiv:2402.03216}, 2024.

\bibitem{emb2024mxbai}
\BIBentryALTinterwordspacing
S.~Lee, A.~Shakir, D.~Koenig, and J.~Lipp. (2024) Open source strikes bread - new fluffy embeddings model. [Online]. Available: \url{https://www.mixedbread.ai/blog/mxbai-embed-large-v1}
\BIBentrySTDinterwordspacing

\bibitem{chambi2016better}
S.~Chambi, D.~Lemire, O.~Kaser, and R.~Godin, ``Better bitmap performance with roaring bitmaps,'' \emph{Software: Practice and Experience}, vol.~46, no.~5, pp. 709--719, 2016.

\bibitem{LOCOMO}
A.~Maharana, D.~Lee, S.~Tulyakov, M.~Bansal, F.~Barbieri, and Y.~Fang, ``Evaluating very long-term conversational memory of {LLM} agents,'' in \emph{Proceedings of the 62nd Annual Meeting of the Association for Computational Linguistics (ACL)}, 2024, pp. 13\,851--13\,870.

\bibitem{hotpotqa}
Z.~Yang, P.~Qi, S.~Zhang, Y.~Bengio, W.~W. Cohen, R.~Salakhutdinov, and C.~D. Manning, ``Hotpotqa: {A} dataset for diverse, explainable multi-hop question answering,'' in \emph{Proceedings of the 2018 Conference on Empirical Methods in Natural Language Processing (EMNLP)}, 2018, pp. 2369--2380.

\bibitem{gutierrez2025rag}
B.~J. Guti{\'e}rrez, Y.~Shu, W.~Qi, S.~Zhou, and Y.~Su, ``From rag to memory: Non-parametric continual learning for large language models,'' in \emph{International Conference on Machine Learning (ICML)}, 2025, pp. 21\,497--21\,515.

\bibitem{gutierrez2024hipporag}
B.~J. Guti{\'e}rrez, Y.~Shu, Y.~Gu, M.~Yasunaga, and Y.~Su, ``Hipporag: Neurobiologically inspired long-term memory for large language models,'' \emph{Advances in neural information processing systems (NeurIPS)}, vol.~37, pp. 59\,532--59\,569, 2024.

\bibitem{LightRAG}
Z.~Guo, L.~Xia, Y.~Yu, T.~Ao, and C.~Huang, ``{LightRAG}: Simple and fast retrieval-augmented generation,'' in \emph{Findings of the Association for Computational Linguistics (EMNLP)}.\hskip 1em plus 0.5em minus 0.4em\relax Association for Computational Linguistics, 2025, pp. 10\,746--10\,761.

\bibitem{ram2019revisiting}
P.~Ram and K.~Sinha, ``Revisiting kd-tree for nearest neighbor search,'' in \emph{Proceedings of the {ACM} {SIGKDD} International Conference on Knowledge Discovery {\&} Data Mining}, 2019, pp. 1378--1388.

\bibitem{DasguptaF08}
S.~Dasgupta and Y.~Freund, ``Random projection trees and low dimensional manifolds,'' in \emph{Proceedings of the Annual {ACM} Symposium on Theory of Computing (STOC)}, 2008, pp. 537--546.

\bibitem{WangWJLZZH14}
J.~Wang, N.~Wang, Y.~Jia, J.~Li, G.~Zeng, H.~Zha, and X.~Hua, ``Trinary-projection trees for approximate nearest neighbor search,'' \emph{{IEEE} Trans. Pattern Anal. Mach. Intell.}, vol.~36, no.~2, pp. 388--403, 2014.

\bibitem{GionisIM99}
A.~Gionis, P.~Indyk, and R.~Motwani, ``Similarity search in high dimensions via hashing,'' in \emph{Proceedings of International Conference on Very Large Data Bases (VLDB)}, 1999, pp. 518--529.

\bibitem{LiZSWT020}
M.~Li, Y.~Zhang, Y.~Sun, W.~Wang, I.~W. Tsang, and X.~Lin, ``{I/O} efficient approximate nearest neighbour search based on learned functions,'' in \emph{{IEEE} International Conference on Data Engineering (ICDE)}, 2020, pp. 289--300.

\bibitem{zheng2020pm}
B.~Zheng, Z.~Xi, L.~Weng, N.~Q.~V. Hung, H.~Liu, and C.~S. Jensen, ``Pm-lsh: A fast and accurate lsh framework for high-dimensional approximate nn search,'' \emph{Proc. {VLDB} Endow.}, vol.~13, no.~5, pp. 643--655, 2020.

\bibitem{huang2015query}
Q.~Huang, J.~Feng, Y.~Zhang, Q.~Fang, and W.~Ng, ``Query-aware locality-sensitive hashing for approximate nearest neighbor search,'' \emph{Proc. {VLDB} Endow.}, vol.~9, no.~1, pp. 1--12, 2015.

\bibitem{zhang2014composite}
T.~Zhang, C.~Du, and J.~Wang, ``Composite quantization for approximate nearest neighbor search,'' in \emph{Proceedings of the International Conference on Machine Learning (ICML)}, vol.~32, 2014, pp. 838--846.

\bibitem{gao2024rabitq}
J.~Gao and C.~Long, ``Rabitq: Quantizing high-dimensional vectors with a theoretical error bound for approximate nearest neighbor search,'' \emph{Proc. {ACM} Manag. Data}, vol.~2, no.~3, p. 167, 2024.

\bibitem{jegou2010product}
H.~J{\'{e}}gou, M.~Douze, and C.~Schmid, ``Product quantization for nearest neighbor search,'' \emph{{IEEE} Trans. Pattern Anal. Mach. Intell.}, vol.~33, no.~1, pp. 117--128, 2011.

\bibitem{ge2013optimized}
T.~Ge, K.~He, Q.~Ke, and J.~Sun, ``Optimized product quantization for approximate nearest neighbor search,'' in \emph{Proceedings of the IEEE Conference on Computer Vision and Pattern Recognition (CVPR)}, 2013, pp. 2946--2953.

\bibitem{malkov2018efficient}
Y.~A. Malkov and D.~A. Yashunin, ``Efficient and robust approximate nearest neighbor search using hierarchical navigable small world graphs,'' \emph{{IEEE} Trans. Pattern Anal. Mach. Intell.}, vol.~42, no.~4, pp. 824--836, 2020.

\bibitem{wang2021comprehensive}
M.~Wang, X.~Xu, Q.~Yue, and Y.~Wang, ``A comprehensive survey and experimental comparison of graph-based approximate nearest neighbor search,'' \emph{Proc. {VLDB} Endow.}, vol.~14, no.~11, pp. 1964--1978, 2021.

\bibitem{fu2017fast}
C.~Fu, C.~Xiang, C.~Wang, and D.~Cai, ``Fast approximate nearest neighbor search with the navigating spreading-out graph,'' \emph{Proc. {VLDB} Endow.}, vol.~12, no.~5, pp. 461--474, 2019.

\bibitem{chen2025maximum}
T.~Chen, C.~Fu, K.~Wang, X.~Ke, Y.~Gao, W.~Zhou, Y.~Ni, and A.~Zeng, ``Maximum inner product is query-scaled nearest neighbor,'' \emph{arXiv:2503.06882}, 2025.

\bibitem{peng2023efficient}
Y.~Peng, B.~Choi, T.~N. Chan, J.~Yang, and J.~Xu, ``Efficient approximate nearest neighbor search in multi-dimensional databases,'' \emph{Proc. {ACM} Manag. Data}, vol.~1, no.~1, pp. 54:1--54:27, 2023.

\bibitem{wang2025boosting}
H.~Wang, W.~Wu, C.~Luo, A.~Bian, C.~Meng, Y.~Wu, and J.~Sun, ``Boosting accuracy and efficiency for vector retrieval with local scaling graph,'' in \emph{IEEE International Conference on Data Engineering (ICDE)}, 2025, pp. 336--348.

\bibitem{fu2025vista}
Y.~Fu, C.~Chen, Y.~Chen, W.-F. Wong, and B.~He, ``Vista: Vector indexing and search for large-scale imbalanced datasets,'' in \emph{2025 IEEE 41st International Conference on Data Engineering (ICDE)}, 2025, pp. 543--556.

\bibitem{li2025fast}
B.~Li, X.~Yan, and S.~Lu, ``Fast-convergent proximity graphs for approximate nearest neighbor search,'' \emph{arXiv:2510.05975}, 2025.

\bibitem{WangWKGZZ25}
M.~Wang, H.~Wu, X.~Ke, Y.~Gao, Y.~Zhu, and W.~Zhou, ``Accelerating graph indexing for {ANNS} on modern cpus,'' \emph{Proc. {ACM} Manag. Data}, vol.~3, no.~3, pp. 123:1--123:29, 2025.

\bibitem{GouGXL25}
Y.~Gou, J.~Gao, Y.~Xu, and C.~Long, ``Symphonyqg: Towards symphonious integration of quantization and graph for approximate nearest neighbor search,'' \emph{Proc. {ACM} Manag. Data}, vol.~3, no.~1, pp. 80:1--80:26, 2025.

\bibitem{gao2023high}
J.~Gao and C.~Long, ``High-dimensional approximate nearest neighbor search: with reliable and efficient distance comparison operations,'' \emph{Proc. {ACM} Manag. Data}, vol.~1, no.~2, pp. 137:1--137:27, 2023.

\bibitem{YangLJZWSJW25}
M.~Yang, W.~Li, J.~Jin, X.~Zhong, X.~Wang, Z.~Shen, W.~Jia, and W.~Wang, ``Effective and general distance computation for approximate nearest neighbor search,'' in \emph{{IEEE} International Conference on Data Engineering (ICDE)}, 2025, pp. 1098--1110.

\bibitem{DengCZWZZ24}
L.~Deng, P.~Chen, X.~Zeng, T.~Wang, Y.~Zhao, and K.~Zheng, ``Efficient data-aware distance comparison operations for high-dimensional approximate nearest neighbor search,'' \emph{Proc. {VLDB} Endow.}, vol.~18, no.~3, pp. 812--821, 2024.

\bibitem{li2020improving}
C.~Li, M.~Zhang, D.~G. Andersen, and Y.~He, ``Improving approximate nearest neighbor search through learned adaptive early termination,'' in \emph{Proceedings of the 2020 International Conference on Management of Data (SIGMOD)}, 2020, pp. 2539--2554.

\bibitem{YangXLYGWPC25}
S.~Yang, J.~Xie, Y.~Liu, J.~X. Yu, X.~Gao, Q.~Wang, Y.~Peng, and J.~Cui, ``Revisiting the index construction of proximity graph-based approximate nearest neighbor search,'' \emph{Proc. {VLDB} Endow.}, vol.~18, no.~6, pp. 1825--1838, 2025.

\bibitem{abs-2508-08744}
Z.~Li, X.~Ke, Y.~Zhu, B.~Yu, B.~Zheng, and Y.~Gao, ``Scalable graph indexing using gpus for approximate nearest neighbor search,'' \emph{arXiv:2508.08744}, 2025.

\bibitem{YueZXXZDGZJ25}
Z.~Yue, B.~Zheng, L.~Xu, K.~Xu, S.~Zhang, Y.~Du, Y.~Gao, X.~Zhou, and C.~S. Jensen, ``Select edges wisely: Monotonic path aware graph layout optimization for disk-based {ANN} search,'' \emph{Proc. {VLDB} Endow.}, vol.~18, no.~11, pp. 4337--4349, 2025.

\bibitem{wang2024starling}
M.~Wang, W.~Xu, X.~Yi, S.~Wu, Z.~Peng, X.~Ke, Y.~Gao, X.~Xu, R.~Guo, and C.~Xie, ``Starling: An i/o-efficient disk-resident graph index framework for high-dimensional vector similarity search on data segment,'' \emph{Proc. {ACM} Manag. Data}, vol.~2, no.~1, pp. 14:1--14:27, 2024.

\bibitem{jayaram2019diskann}
S.~J. Subramanya, Devvrit, H.~V. Simhadri, R.~Krishnaswamy, and R.~Kadekodi, ``Diskann: Fast accurate billion-point nearest neighbor search on a single node,'' in \emph{Advances in Neural Information Processing Systems (NeurIPS)}, 2019, pp. 13\,748--13\,758.

\bibitem{LiuZYRZJ25}
D.~Liu, B.~Zheng, Z.~Yue, F.~Ruan, X.~Zhou, and C.~S. Jensen, ``Wolverine: Highly efficient monotonic search path repair for graph-based {ANN} index updates,'' \emph{Proc. {VLDB} Endow.}, vol.~18, no.~7, pp. 2268--2280, 2025.

\bibitem{MaZZCGG24}
R.~Ma, Y.~Zhu, B.~Zheng, L.~Chen, C.~Ge, and Y.~Gao, ``{GTI:} graph-based tree index with logarithm updates for nearest neighbor search in high-dimensional spaces,'' \emph{Proc. {VLDB} Endow.}, vol.~18, no.~4, pp. 986--999, 2024.

\bibitem{abs-2502-13826}
H.~Xu, M.~D. Manohar, P.~A. Bernstein, B.~Chandramouli, R.~Wen, and H.~V. Simhadri, ``In-place updates of a graph index for streaming approximate nearest neighbor search,'' \emph{arXiv:2502.13826}, 2025.

\bibitem{ye2025compass}
C.~Ye, X.~Yan, and E.~Lo, ``Compass: General filtered search across vector and structured data,'' \emph{arXiv:2510.27141}, 2025.

\bibitem{NaviX}
G.~Sehgal and S.~Salihoglu, ``Navix: {A} native vector index design for graph dbmss with robust predicate-agnostic search performance,'' \emph{Proc. {VLDB} Endow.}, vol.~18, no.~11, pp. 4438--4450, 2025.

\bibitem{ACORN}
L.~Patel, P.~Kraft, C.~Guestrin, and M.~Zaharia, ``{ACORN:} performant and predicate-agnostic search over vector embeddings and structured data,'' \emph{Proc. {ACM} Manag. Data}, vol.~2, no.~3, p. 120, 2024.

\bibitem{SIEVE}
Z.~Li, S.~Huang, W.~Ding, Y.~Park, and J.~Chen, ``{SIEVE:} effective filtered vector search with collection of indexes,'' \emph{Proc. {VLDB} Endow.}, vol.~18, no.~11, pp. 4723--4736, 2025.

\bibitem{abs-2508-16263}
M.~Li, X.~Yan, B.~Lu, Y.~Zhang, J.~Cheng, and C.~Ma, ``Attribute filtering in approximate nearest neighbor search: An in-depth experimental study,'' \emph{arXiv:2508.16263}, 2025.

\bibitem{abs-2507-21989}
P.~Iff, P.~Bruegger, M.~Chrapek, M.~Besta, and T.~Hoefler, ``Benchmarking filtered approximate nearest neighbor search algorithms on transformer-based embedding vectors,'' \emph{arXiv:2507.21989}, 2025.

\bibitem{abs-2505-06501}
Y.~Lin, K.~Zhang, Z.~He, Y.~Jing, and X.~S. Wang, ``Survey of filtered approximate nearest neighbor search over the vector-scalar hybrid data,'' \emph{arXiv:2505.06501}, 2025.

\bibitem{abs-2509-07789}
J.~Shi, Y.~Cai, and W.~Zheng, ``Filtered approximate nearest neighbor search: {A} unified benchmark and systematic experimental study [experiment, analysis {\&} benchmark],'' \emph{arXiv:2509.07789}, 2025.

\bibitem{LuoQZD25}
J.~Luo, M.~Qiao, C.~Zuo, and D.~Deng, ``Tag-filtered approximate nearest neighbor search,'' in \emph{{IEEE} International Conference on Data Engineering (ICDE)}, 2025, pp. 3642--3654.

\bibitem{GollapudiKSKBRL23}
S.~Gollapudi, N.~Karia, V.~Sivashankar, R.~Krishnaswamy, N.~Begwani, S.~Raz, Y.~Lin, Y.~Zhang, N.~Mahapatro, P.~Srinivasan, A.~Singh, and H.~V. Simhadri, ``Filtered-diskann: Graph algorithms for approximate nearest neighbor search with filters,'' in \emph{Proceedings of the {ACM} Web Conference (WWW)}, 2023, pp. 3406--3416.

\bibitem{LiangZYCSC24}
A.~Liang, P.~Zhang, B.~Yao, Z.~Chen, Y.~Song, and G.~Cheng, ``{UNIFY:} unified index for range filtered approximate nearest neighbors search,'' \emph{Proc. {VLDB} Endow.}, vol.~18, no.~4, pp. 1118--1130, 2024.

\bibitem{WangHTZZ25}
Y.~Wang, Z.~He, Y.~Tong, Z.~Zhou, and Y.~Zhong, ``Timestamp approximate nearest neighbor search over high-dimensional vector data,'' in \emph{{IEEE} International Conference on Data Engineering (ICDE)}, 2025, pp. 3043--3055.

\bibitem{abs-2508-18617}
Z.~Wang, J.~Zhang, and W.~Hu, ``Wow: {A} window-to-window incremental index for range-filtering approximate nearest neighbor search,'' \emph{arXiv:2508.18617}, 2025.

\bibitem{ZuoQZLD24}
C.~Zuo, M.~Qiao, W.~Zhou, F.~Li, and D.~Deng, ``Serf: Segment graph for range-filtering approximate nearest neighbor search,'' \emph{Proc. {ACM} Manag. Data}, vol.~2, no.~1, pp. 69:1--69:26, 2024.

\bibitem{CaiSCZ24}
Y.~Cai, J.~Shi, Y.~Chen, and W.~Zheng, ``Navigating labels and vectors: {A} unified approach to filtered approximate nearest neighbor search,'' \emph{Proc. {ACM} Manag. Data}, vol.~2, no.~6, pp. 246:1--246:27, 2024.

\bibitem{HQANN}
W.~Wu, J.~He, Y.~Qiao, G.~Fu, L.~Liu, and J.~Yu, ``{HQANN:} efficient and robust similarity search for hybrid queries with structured and unstructured constraints,'' in \emph{Proceedings of the {ACM} International Conference on Information {\&} Knowledge Management (CIKM)}, 2022, pp. 4580--4584.

\bibitem{NHQ}
M.~Wang, L.~Lv, X.~Xu, Y.~Wang, Q.~Yue, and J.~Ni, ``An efficient and robust framework for approximate nearest neighbor search with attribute constraint,'' in \emph{Advances in Neural Information Processing Systems (NeurIPS)}, 2023.

\bibitem{abs-2509-19767}
A.~Heidari, W.~Zhang, and Y.~Xiong, ``Fusedann: Convexified hybrid {ANN} via attribute-vector fusion,'' \emph{arXiv:2509.19767}, 2025.

\bibitem{abs-2508-01405}
M.~Wang, B.~Tan, Y.~Gao, H.~Jin, Y.~Zhang, X.~Ke, X.~Xu, and Y.~Zhu, ``Balancing the blend: An experimental analysis of trade-offs in hybrid search,'' \emph{arXiv:2508.01405}, 2025.

\bibitem{BruchNIL24a}
S.~Bruch, F.~M. Nardini, A.~Ingber, and E.~Liberty, ``Bridging dense and sparse maximum inner product search,'' \emph{{ACM} Trans. Inf. Syst.}, vol.~42, no.~6, pp. 151:1--151:38, 2024.

\bibitem{LuanETC21}
Y.~Luan, J.~Eisenstein, K.~Toutanova, and M.~Collins, ``Sparse, dense, and attentional representations for text retrieval,'' \emph{Trans. Assoc. Comput. Linguistics}, vol.~9, pp. 329--345, 2021.

\bibitem{YangCHQY24}
Y.~Yang, P.~Carlson, S.~He, Y.~Qiao, and T.~Yang, ``Cluster-based partial dense retrieval fused with sparse text retrieval,'' in \emph{Proceedings of the International {ACM} {SIGIR} Conference on Research and Development in Information Retrieval}, 2024, pp. 2327--2331.

\bibitem{Blend-RAG}
K.~Sawarkar, A.~Mangal, and S.~R. Solanki, ``Blended {RAG:} improving {RAG} (retriever-augmented generation) accuracy with semantic search and hybrid query-based retrievers,'' in \emph{{IEEE} International Conference on Multimedia Information Processing and Retrieval (MIPR)}, 2024, pp. 155--161.

\bibitem{li2025all}
Z.~Li, Y.~Li, Y.~Zhu, Z.~Chen, and Y.~Gao, ``All-in-one graph-based indexing for hybrid search on gpus,'' \emph{arXiv:2511.00855}, 2025.

\bibitem{ChenZLBN22}
T.~Chen, M.~Zhang, J.~Lu, M.~Bendersky, and M.~Najork, ``Out-of-domain semantics to the rescue! zero-shot hybrid retrieval models,'' in \emph{Advances in Information Retrieval - European Conference on {IR} Research (ECIR)}, vol. 13185, 2022, pp. 95--110.

\bibitem{ragflow}
``Ragflow,'' \url{https://github.com/infiniflow/ragflow}, 2025, [Online; accessed 23-September-2025].

\end{thebibliography}

\clearpage
\end{document}